\documentclass[preprint]{aastex}

\shorttitle{GRMHD Accretion Tori Simulations}
\shortauthors{De Villiers and Hawley}

\begin{document}

\title{Global General Relativistic Magnetohydrodynamic Simulations of 
Accretion Tori}


\author{Jean-Pierre De Villiers and John F. Hawley}
\affil{Astronomy Department\\
University of Virginia\\ 
P.O. Box 3818, University Station\\
Charlottesville, VA 22903-0818}
\email{jd5v@virginia.edu; jh8h@virginia.edu}

\begin{abstract}

This paper presents an initial survey of the properties of accretion
flows in the Kerr metric from three-dimensional, general relativistic
magnetohydrodynamic simulations of accretion tori.  We consider three
fiducial models of tori around rotating, both prograde and retrograde,
and nonrotating black holes; these three fiducial models are also
contrasted with axisymmetric simulations and a pseudo-Newtonian
simulation with equivalent initial conditions to delineate the
limitations of these approximations.  There are both qualitative and
quantitative differences in the fiducial models, with many of these
effects attributable to the location of the marginally stable orbit,
$r_{ms}(a)$, both with respect to the initial torus and in absolute
terms.  In the retrograde model, the initial inner edge of the torus is
close to $r_{ms}$ and little angular momentum need be lost to drive
accretion, whereas in the prograde case the gas must slowly accrete over
a significant distance and shed considerable angular momentum.
Evolution is driven by the magneto-rotational instability and the
nonzero Maxwell stresses produced by the turbulence, and results in a
redistribution of the specific angular momentum to near-Keplerian
values.  The magnetic energy remains subthermal within the turbulent
disk, but dominates in the final plunging flow into the hole.  The
Maxwell stress remains nonzero in this plunging flow inside of $r_{ms}$
and the fluid's specific angular momentum continues to drop.  The
accretion rate into the hole is highly time variable and is determined
by the rate at which gas from the turbulent disk is fed into the
plunging flow past $r_{ms}$.  The retrograde model, with the largest
$r_{ms}$, shows the least variability in accretion rate.  While
accretion variability is a function of $a$, the turbulence itself is
also intrinsically variable.  A magnetized backflowing corona and an
evacuated, magnetized funnel are features of all models.

\end{abstract}


\keywords{Black holes - magnetohydrodynamics - instabilities - stars:accretion}

\section{Introduction}

Numerical experiments are playing an increasingly central role in
investigating accretion into compact objects.  A particularly important
application is accretion into black holes, either solar-mass black
holes in X-ray binaries, or supermassive black holes in active galactic
nuclei, quasars, and the like.  The investigation of such systems
requires the development of algorithms and simulation codes that solve
the equations of magnetohydrodynamics (MHD) in full general relativity
(GR).  To date, however, the only global three-dimensional MHD black
hole accretion simulations have been done using a pseudo-Newtonian
potential.  The pseudo-Newtonian potential is $\Phi = - GM/(r-r_g)$,
where the gravitational radius, $r_g$, is equated with the
Schwarzschild radius (Paczy\'nski \& Wiita 1980).  This potential
reproduces several important aspects of the Schwarzschild potential, in
particular the qualitative behavior associated with the innermost
stable orbit.  But the system is still fundamentally Newtonian, and
cannot account for the effects of a rotating black hole.  Fully
relativistic simulations are necessary, and in this paper we present
the first such three-dimensional MHD black hole accretion models.

Since the pseudo-Newtonian model is substantially simpler than full
relativity, it has been the natural starting point for accretion
simulations.  We begin by briefly reviewing some of those results.
Hawley (2000; hereafter H00) computed fully global, three-dimensional
prototype models of magnetized accretion tori using Newtonian MHD.  H00
compared a variety of approximations:  the ``cylindrical disk'' limit
(in which the vertical component of gravity is ignored),
two-dimensional versus three-dimensional simulations, and Newtonian
versus pseudo-Newtonian potentials.  This work demonstrated in a global
context that the magnetorotational instability (MRI; Balbus \& Hawley
1991) drives the evolution of accretion disks by generating turbulence
and transporting angular momentum through significant Maxwell
stresses.  Tori with near-constant angular momentum distributions,
which served as initial conditions, are particularly unstable, and
quickly evolve to near-Keplerian disks.  It was further found that the
Maxwell stress did not go to zero at the radius of the marginally
stable orbit, $r_{ms}$, the point traditionally regarded as the
effective inner boundary of the disk.  Instead, the specific angular
momentum of the gas continued to decline inside $r_{ms}$.

Questions of large-scale disk structure and evolution were addressed
with global MHD simulations by Hawley, Balbus \& Stone (2001) and
Hawley \& Balbus (2002).  This work examined the accretion of gas from
a torus at $100\,r_g$, and the subsequent formation of a disk interior
to that point.  In this model the accretion flow was assumed to be
nonradiating.   As a consequence, the resulting disk was vertically
thick,  with an angular momentum distribution that had a Keplerian
slope.  The flow also featured a substantial outward flowing magnetized
corona.  Further, the accretion rate into the near hole region was
larger than the accretion rate through the equatorial opening in the
centrifugal barrier.  Gas ``backed up'' at this point, creating an hot
torus in the inner region near the hole, and an outflow along the
centrifugal wall surrounding the evacuated funnel.  The primary
elements of this model---thick, nearly Keplerian disk, magnetized
coronal backflow, hot inner torus---appear to be generic features of
nonradiative accretion flows.

The properties of the inner region of an accretion disk are worth
particularly close scrutiny since the most extreme relativistic effects
and the greatest energy release happen near $r_{ms}$.  The properties of
the inner edge were the focus of the work of Hawley \& Krolik (2001,
2002) which used relatively high resolution simulations to study the
inflow past $r_{ms}$ and into the black hole.  Cylindrical disk studies
investigating this region were also done by Armitage, Reynolds, \&
Chiang (2001).  Krolik \& Hawley (2002) examined the implications of the
simulations for the inner edge of a black hole accretion disk.  The
essential points are as follows:  an accretion flow must transition from
turbulence within the disk to a plunging infall into the hole, and
simulations find that the location of this transition is highly
time-variable.  Generally the transition begins outside of $r_{ms}$, but
the full transonic plunging inflow is not fully established until some
point inside of $r_{ms}$.  The inner disk edge is not abrupt:  average
flow variables, including the stress, are relatively smooth, albeit
time-varying, functions of radius.  Magnetic stress continues to be
important even within the plunging region.  In fact, the magnetic stress
becomes proportionally more important as flux-freezing increases the
field strength relative to the gas pressure, and increases the
correlation between the radial and azimuthal field components.

As remarked, all the above work was done within the context of a
pseudo-Newtonian potential.  Just how well does that potential
model the most important behavior in accretion flows?  When is that
approximation sufficient and when does it fail?  What unique
properties in the accretion flow are created by the spin of the hole
and frame dragging, and do these produce unique observational
signatures?  Answering questions such as these requires a fully
relativistic treatment.

The development of new codes for general relativistic MHD is ongoing.
Recent work includes that of Koide, Shibata \& Kudoh (1999), Gammie,
McKinney, \& T\'oth (2003), and our own effort (De Villiers \& Hawley
2003; hereafter DH03).  In DH03 we describe the development and testing
of the general relativistic MHD code used in this paper.  The
hydrodynamic portion of this code has already been employed (De
Villiers \& Hawley 2002; hereafter DH02) to study the evolution of
accretion tori that are subject to the Papaloizou-Pringle instability
(Papaloizou \& Pringle 1984) in the full Kerr metric.  Here we turn our
attention to the first in a series of global simulations of accretion
tori in the Kerr metric using the equations of MHD.  Our aim with these
models is to lay the groundwork for future, more detailed, black hole
accretion simulations.

In this paper we examine a relativistic version of the constant angular
momentum torus model GT1 studied in H00.  In \S2 we write the equations
of general relativistic MHD in the form that the code uses.  We
describe the constant angular momentum tori that serve as initial
conditions for the models in this paper.  We list a number of the
physics diagnostics used in the simulations.  In \S3 we present the
results of the three-dimensional global MHD simulations, specifically a
Schwarzschild hole, a prograde, and a retrograde Kerr hole with near
maximal spin.  We contrast three-dimensional simulations with
two-dimensional axisymmetric models.  We also compare a simulation that
includes the full range in $\phi$ ($0 \le \phi \le 2 \pi$) with the
results from the more economical quarter-plane simulation ($0 \le \phi
\le \pi/2$), to examine the emerging impression that quarter-plane
simulations give qualitatively the same results, and are adequate for
many studies (Hawley 2001; Nelson \& Papaloizou 2003).  We contrast the
results from the Schwarzschild simulation with an equivalent
pseudo-Newtonian model.  In \S4 we review and summarize our findings.

\section{Equations of General Relativistic Magnetohydrodynamics}

We are studying the evolution of a magnetized fluid in the background
spacetime of a Kerr (rotating) black hole.  We adopt the familiar
Boyer-Lindquist coordinates, $(t,r,\theta,\phi)$, for which the line
element has the form,
\begin{equation}\label{kerr}
{ds}^2=g_{t t}\,{dt}^2+2\,g_{t \phi}\,{dt}\,{d \phi}+g_{r r}\,{dr}^2+
 g_{\theta \theta}\,{d \theta}^2+g_{\phi \phi}\,{d \phi}^2 .
\end{equation}
We use the metric signature $(-,+,+,+)$, along with geometrodynamic
units where $G = c = 1$.  Time and distance are in units of 
the black hole mass, $M$.  The
determinant of the 4-metric is $g$, and $\sqrt{-g} =
\alpha\,\sqrt{\gamma}$, where $\alpha$ is the lapse function,
$\alpha=1/\sqrt{-g^{tt}}$, and $\gamma$ is the determinant of the
spatial $3$-metric.

The equations of general relativistic ideal MHD are the equation of
baryon conservation, the conservation of the fluid energy-momentum
tensor, and the induction equation.  While the equations themselves are
determined, there are a large number of ways that they can be expressed
and solved numerically.  In our methods paper (DH03) we derive useful
forms for these equations, along with a set of primary and secondary
variables especially suited to numerical evolution.   The first of
these is the equation of mass conservation,
\begin{equation}\label{masscons}
\partial_t\,D + {1 \over
\sqrt{\gamma}}\,\partial_j\,(D\,\sqrt{\gamma}\,V^j)  = 0,
\end{equation}
where $D=\rho\,W$ is the auxiliary density variable, $\rho$ the density,
and $V^j = U^j/U^t$ is the transport velocity, $U^\mu$ is the 
four-velocity,  $W$ is the relativistic gamma-factor, and 
$U^t = W/\alpha$. Spatial indices are indicated by roman characters 
$i,j=1,2,3$.

The internal energy equation,
\begin{equation}\label{encons}
\partial_t\,E + {1 \over
\sqrt{\gamma}}\,\partial_j\,(E\,\sqrt{\gamma}\,V^j) + P\,\partial_t\,W
+ {P \over \sqrt{\gamma}}\,\partial_j\,(W\,\sqrt{\gamma}\,V^j) =  0,
\end{equation}
evolves $E= \epsilon\,D$, the
auxiliary energy variable, where $\epsilon$ is the specific internal
energy.   The isotropic pressure $P$ is determined by the equation
of state of an ideal gas, $P=\rho\,\epsilon\,(\Gamma-1)$, where
$\Gamma$ is the adiabatic exponent.  For these simulations we take
$\Gamma=5/3$.

The equation of momentum conservation, 
\begin{equation}\label{momcons}
\partial_t\left(S_j-\alpha\,b_j\,b^t\right)+
  {1 \over \sqrt{\gamma}}\,
  \partial_i\,\sqrt{\gamma}\,\left(S_j\,V^i-\alpha\,b_j\,b^i\right)+
  {1 \over 2}\,\left({S_\epsilon\,S_\mu \over S^t}-
  \alpha\,b_\mu\,b_\epsilon\right)\,
  \partial_j\,g^{\mu\,\epsilon}+
  \alpha\,\partial_j\left(P+{{\|b\|}^2 \over 2}\right) = 0 .
\end{equation}
is written in terms of the auxillary four-momentum,
$S_\mu = (\rho\,h\ + {\|b\|}^2)\,W\,U_\mu$, where $h=1 +
\epsilon + P/\rho$ is the relativistic enthalpy, 
$b^\mu$ is the magnetic field four-vector in the
rest-frame of the fluid, and ${\|b\|}^2=g^{\mu\,\nu}\,b_\mu\,b_\nu$.
The momentum is subject to the  normalization condition
$g^{\mu \nu}\,S_\mu\,S_\nu = -{(\rho\,h+{\|b\|}^2)}^2\,W^2$,
which is algebraically equivalent to the more familiar velocity
normalization $U^\mu\,U_\mu=-1$.

Finally, the induction equation  is solved in the form
\begin{equation}
F_{\alpha \beta , \gamma} +
F_{\beta  \gamma, \alpha} +
F_{\gamma \alpha, \beta } = 0,
\end{equation}
where $F_{\alpha\beta}$ is the electromagnetic field strength tensor,
written in terms of the constrained transport (CT) magnetic field
variables of Evans \& Hawley (1988),
\begin{equation}
{\cal{B}}^r      = F_{\phi \theta} \, , \,
{\cal{B}}^\theta = F_{r \phi} \, , \,
{\cal{B}}^\phi   = F_{\theta r} .
\end{equation}
We assume infinite conductivity (the
flux-freezing condition), $F^{\mu \nu}\,U_\nu = 0$.
The induction equation then reads
\begin{eqnarray}\label{ct}
\partial_j \left({\cal{B}}^j\right) & = 0 & (\nu=0) ,\\
\label{ct.3b}
\partial_t \left({\cal{B}}^i\right) -
\partial_j \left(V^i\,{\cal{B}}^j-{\cal{B}}^i\,V^j\right)& = 0 & (\nu=i),
\end{eqnarray}
where $V^\mu = U^\mu/U^t$ is the transport velocity with $U^t = W/\alpha$.
The CT variables ${\cal{B}}^i$ are related to the magnetic four-vector
$b^\mu$ by
\begin{equation}
\label{fmunudef.3}
b^i={{\cal{B}}^i \over \sqrt{4\,\pi}\,\sqrt{\gamma}\,W}+V^i\,b^t,
\end{equation}
and
\begin{eqnarray}\label{bdef.5}
b^t & = & {W \over \sqrt{4 \pi} \alpha^2\,\sqrt{\gamma}}\, 
 \left[g_{rr}\,V^r\,{\cal{B}}^r+
  g_{\theta \theta}\,V^\theta\,{\cal{B}}^\theta+\left(
  g_{\phi \phi}\,V^\phi + g_{t \phi}\right)\,{\cal{B}}^\phi\right] .
\end{eqnarray}
Note that the factor $\sqrt{4\pi}$ is incorporated into the definition
of $b^\mu$.

For reference, the energy-momentum tensor, written 
using the primitive variables, has the familiar form
\begin{equation} \label{tmndef}
{T}^{\mu\,\nu} = \left(\rho\,h\,+ {\|b\|}^2\right){U}^{\mu}\,{U}^{\nu}+
 \left(P +  {{\|b\|}^2 \over 2}\right)\,{g}^{\mu\,\nu}-
 b^\mu\,b^\nu ,
\end{equation}
which readily identifies the magnetic pressure, $P_{mag}={\|b\|}^2/2$.

The GRMHD code evolves time-explicit, operator-split, finite difference
forms of equations (\ref{masscons})---(\ref{momcons}) and evolves the
magnetic field induction equation (\ref{ct.3b}) using the CT algorithm
and a modified method of characteristic technique as described in
DH03.  In this paper we evolve a fiducial constant-angular momentum
torus containing poloidal field loops orbiting around a prograde
rotating Kerr hole, a retrograde rotating Kerr hole, and a nonrotating
Schwarzschild hole.  The fiducial models use a $128\times 128 \times
32$ grid in $(r,\theta, \phi)$.  The azimuthal grid spans the quarter
plane, i.e., $0 \le \phi \le \pi/2$, with periodic boundary conditions
in $\phi$.  One test simulation was run with a full $2\pi$ azimuthal
grid, using $128 \times 128 \times 128$ zones.  The outer radial
boundary is set to $r_{out}=120 M$ in all cases; the inner radial
boundary is located just outside the horizon, with the specific value
depending on the location of the horizon as determined by the Kerr spin
parameter $a$.  The radial grid is spaced logarithmically to maximize
the resolution near the inner boundary.  The radial boundary conditions
are zero gradient boundary conditions, where the contents of the active
zones adjacent to the boundary are copied into neighboring ghost
zones.  The $\theta$-grid runs over $0.02 \pi \le \theta \le 0.98 \pi$,
with reflecting boundary conditions enforced at the polar axis.  This
avoids the substantial reduction in timestep that occurs if the
$\theta$ grid runs up to the coordinate singularity at the axis.  Test
simulations have found no significant differences resulting from the
reduced $\theta$ grid.  Because the accreting gas has angular momentum,
it is excluded from the region near the rotation axis.

A density and energy floor is employed to ensure that the dynamic range
between the vacuum and the accreting gas does not become too great.
The floor is set to be $10^{-7}$ the initial density maximum in the torus.
The timestep is set to one half the minimum time required
for light to cross one grid zone.  Because the grid is fixed, the
timestep too is fixed throughout the simulation.

\subsection{Torus Initial State}

For the first simulations we choose a particular initial state that has
been studied extensively in previous pseudo-Newtonian simulations:  the
constant-angular momentum torus containing weak poloidal magnetic field
loops.  We adopt the definition of specific angular momentum $l
=-U_\phi/U_t$.  The initial axisymmetric torus is generated from the
constant-$l$ axisymmetric GR hydrodynamic equilibrium equations for an
adiabatic gas, as described in Hawley, Smarr, \& Wilson (1984; see also
DH02).  A particular torus is determined by the choice of $l$, which
defines a set of effective potential surfaces,  and the specific
binding energy of the effective potential corresponding to the torus
surface, designated $\left( U_t \right)_s$.  Alternately, one can
specify the radius of the pressure maximum (where $l$ is equal to the
Keplerian value), and the radius of the inner edge of the torus (which
sets $\left( U_t \right)_s$).

The initial magnetic field is obtained from the definition of $F_{\mu
\nu}$ in terms of the $4$-vector potential, $A_\mu$,
\begin{equation}\label{avec.1}
F_{\mu \nu} = \partial_\mu A_{\nu} - \partial_\nu A_{\mu} .
\end{equation}
Our initial field consists of axisymmetric poloidal field loops, laid
down along isodensity surfaces within the torus by defining
$A_{\mu} = (A_t,0,0,A_\phi)$, where
\begin{equation}\label{vecpot}
A_\phi = 
\cases{
k (\rho-\rho_{cut}) & for $\rho \ge \rho_{cut}$ \cr
0 & for $\rho < \rho_{cut}$},
\end{equation}
where $\rho_{cut}$ is a cutoff density corresponding to a particular
isodensity surface within the torus.  Using (\ref{avec.1}), it follows
that ${\cal{B}}^r = -\partial_\theta A_{\phi}$ and ${\cal{B}}^\theta =
\partial_r A_{\phi}$.  The constant $k$ is set by the input parameter
$\beta$, the ratio of the gas pressure to the magnetic pressure, using
the volume-integrated gas pressure divided by the volume-integrated
magnetic energy density in the initial torus.  In all calculations
reported here the average initial field strength is $\beta = 100$.  The
constant $\rho_{cut}$ is chosen to keep the initial magnetic field away
from the outer edge of the disk.  Here we use $\rho_{cut} = 0.5
\rho_{max}$, where $\rho_{max}$ is the maximum density at the center of
the torus.  This choice means that the initial field loops will be
confined well inside the torus.

\subsection{Evolution Diagnostics}

One of the challenges of global, three-dimensional simulations of
complex phenomena is to extract and distill useful diagnostic
information that accurately characterizes the most important physical
properties of the system.  One computes, of course, a complete set of
variables, at all grid zones for all timesteps; choosing from this data
what to evaluate and to save is the issue.  Given the novelty of
three-dimensional MHD GR accretion simulations, there is as yet no
definitive set of diagnostic quantities.  We anticipate that standards
will develop and evolve as our understanding of the simulation dynamics
of black hole accretion flows improves.  In previous pseudo-Newtonian
calculations (e.g. H00), the evolution was characterized by several
reduced quantities extracted from the simulation data.  The set of
diagnostics presented here is larger than those originally discussed in
H00, and serves a first attempt at creating a new set of reduced
quantities in a relativistic framework.

Entire restart files are saved at periodic intervals, and these can
be examined in detail after a simulation.  The more difficult choice is
what data to save at frequent time intervals as a running history of
the simulation.  We have used full dumps of individual variables, both
fundamental and derived; for the present simulations we saved
density, and magnetic and thermal pressure 30 times per orbit.
We also computed a more extensive set of azimuthally-averaged
variables, variables averaged on spherical shells, and
volume-integrated quantities.  We define the average of a quantity
${\cal X}$ on a shell at radius $r$ as
\begin{equation}\label{avgdef}
\langle{\cal X}\rangle(r) = {1 \over {{\cal A}}(r)} \int\int{ 
 {\cal X}\,\sqrt{-g}\, d \theta\,d \phi}
\end{equation}
where the area of a shell is ${\cal{A}}(r)$ and the bounds of integration 
range over the $\theta$ and $\phi$ grids.  
For these simulations we compute shell-averaged values of density,
$\langle\rho\rangle$, angular momentum,
$\langle \rho\, l\rangle$, gas pressure, $\langle P\rangle$,  and
magnetic pressure $\langle{\|b\|}^2\rangle$.  
From these quantities,
we derive quantities such as the density-weighted average 
specific angular momentum,
$\langle l\rangle =\langle \rho\, l \rangle/\langle\rho\rangle$. 

Fluxes through the shell are computed in the same manner, but are
not normalized with the area.  We evaluate the rest mass flux
$\langle\rho\,U^r\rangle$, energy flux 
\begin{equation}
\langle{T^r}_t\rangle = \rho\,U^r\,U_{t}
+{\|b\|}^2\,U^r\,U_{t}-b^r\,b_t ,
\end{equation}
and the angular momentum flux $\langle {T^r}_{\phi}\rangle$,
saved as two separate parts, i.e., the fluid part
\begin{equation}
\langle{T^r}_{\phi\,\rm{(F)}}\rangle = \rho\,U^r\,U_{\phi},
\end{equation}
and the magnetic part,
\begin{equation}
\langle{T^r}_{\phi\,\rm{(M)}}\rangle =
{\|b\|}^2\,U^r\,U_{\phi}-b^r\,b_\phi.
\end{equation}
Again, various quantities can be subsequently derived from these
fluxes and shell averages.

Volume-integrated quantities are computed using
\begin{equation}\label{3avgdef}
\left[{\cal Q}\right] = \int\int\int{ 
 {\cal Q}\,\sqrt{-g}\, dr\,d \theta\,d \phi}.
\end{equation}
The volume-integrated quantities are the total rest mass, 
$\left[ \rho U^t \right]$,  and
total energy $\left[ {T}^{tt}\right]$. 
In all the simulations these integrated values are computed and saved
$30$ times per orbit at the initial pressure maximum.

\section{Results}

\subsection{Retrograde, Schwarzschild, and Prograde Black Holes}

Our fiducial simulations consider and contrast three initial torus
models:  a torus orbiting a Schwarzschild black hole, and tori orbiting
rapidly rotating retrograde and prograde Kerr holes.  Although the
initial conditions cannot be made identical, we choose parameters for
the initial tori that keep the inner edge of the disk and the location
of the initial pressure maximum roughly the same as the black hole
rotation parameter is varied.  Table \ref{params} lists the general
properties of the models, where $l$ denotes the specific angular
momentum, $r_{in}$ the inner edge of the disk (in the equatorial plane),
$(U_{t})_s$ is the binding energy of the surface of the torus,
$r_{Pmax}$ the location of the pressure maximum (also in the equatorial
plane), $T_{orb}$ the orbital period at the pressure maximum in units of
$M$.  For reference we also list $r_{ms}$, the location of the
marginally stable orbit, the values of $(U_\phi )_{ms}$ and $(U_t)_{ms}$
for a particle in a circular orbit at $r_{ms}$ (Bardeen, Press, \&
Teukolsky 1972), and $T_{orb\,(ms)}$, the orbital period at the
marginally stable orbit in units of $M$.  All radii in the table are in
units of $M$.  The retrograde SFR model is an extreme Kerr hole,
$a=-0.998$, while the prograde torus orbits a rapidly rotating Kerr hole
with the rotation parameter $a= 0.9$ chosen so that the marginally
stable orbit is outside the static limit.

\begin{table}[ht]
\caption{\label{params}Global Torus Simulation Parameters.}
\begin{tabular}{lcrrrrcrrrr}
 & & & & & & & & &\\
\hline
Model & a & $l$ & $\left(U_t\right)_s$ & $r_{in}$ & $r_{Pmax}$ & $T_{orb}$ & 
$r_{ms}$ & $\left( U_\phi\right)_{ms}$ & $\left( U_t\right)_{ms}$ & 
$T_{orb\,(ms)}$\\
\hline
\hline
SF0 & 0.000 & 4.50 & -0.980 & 9.5 & 15.3 & 376 & 6.00 & 3.464 & -0.942
& 97.95 \\
SFP & 0.900 & 4.30 & -0.980 & 9.5 & 15.4 & 386 & 2.32 & 2.100 & -0.844
& 31.04 \\
SFR & 0.998 &-4.80 & -0.980 & 9.5 & 15.8 & 388 & 8.99 &-4.233 & -0.962
& 170.0 \\
\hline
\end{tabular}
\end{table}

\begin{figure}[ht] 
\epsscale{0.3}
   \plotone{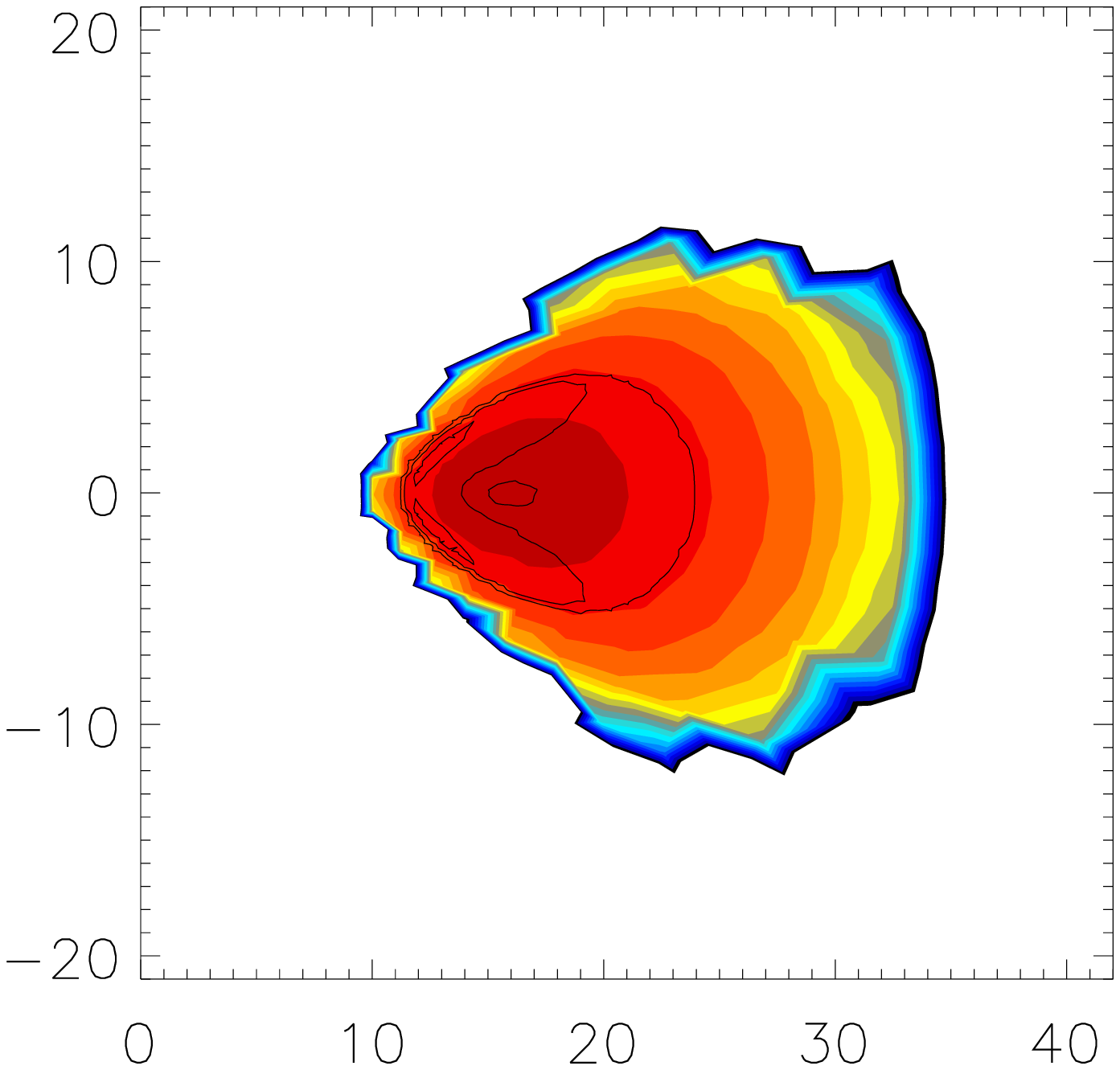}
   \plotone{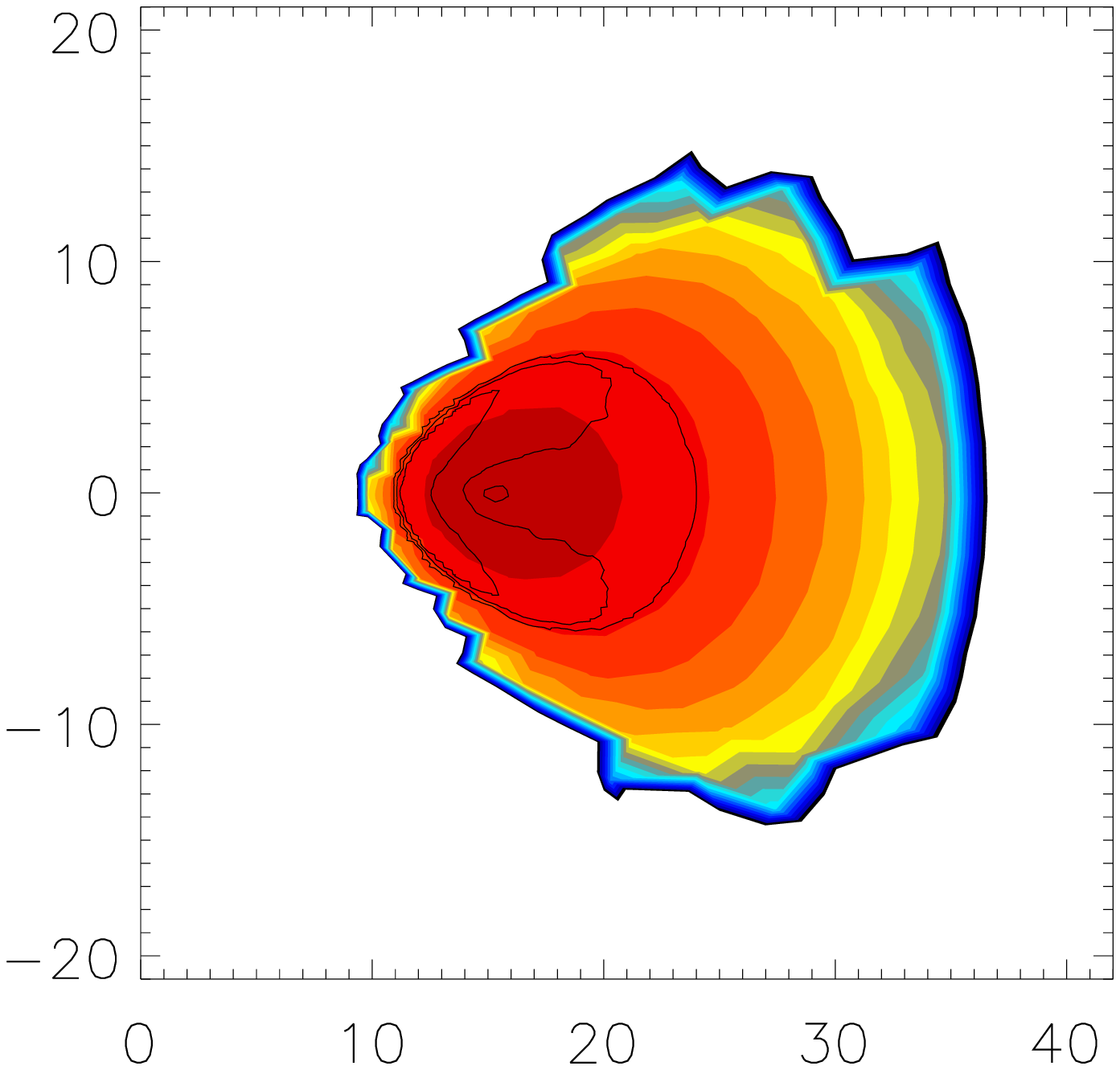}
   \plotone{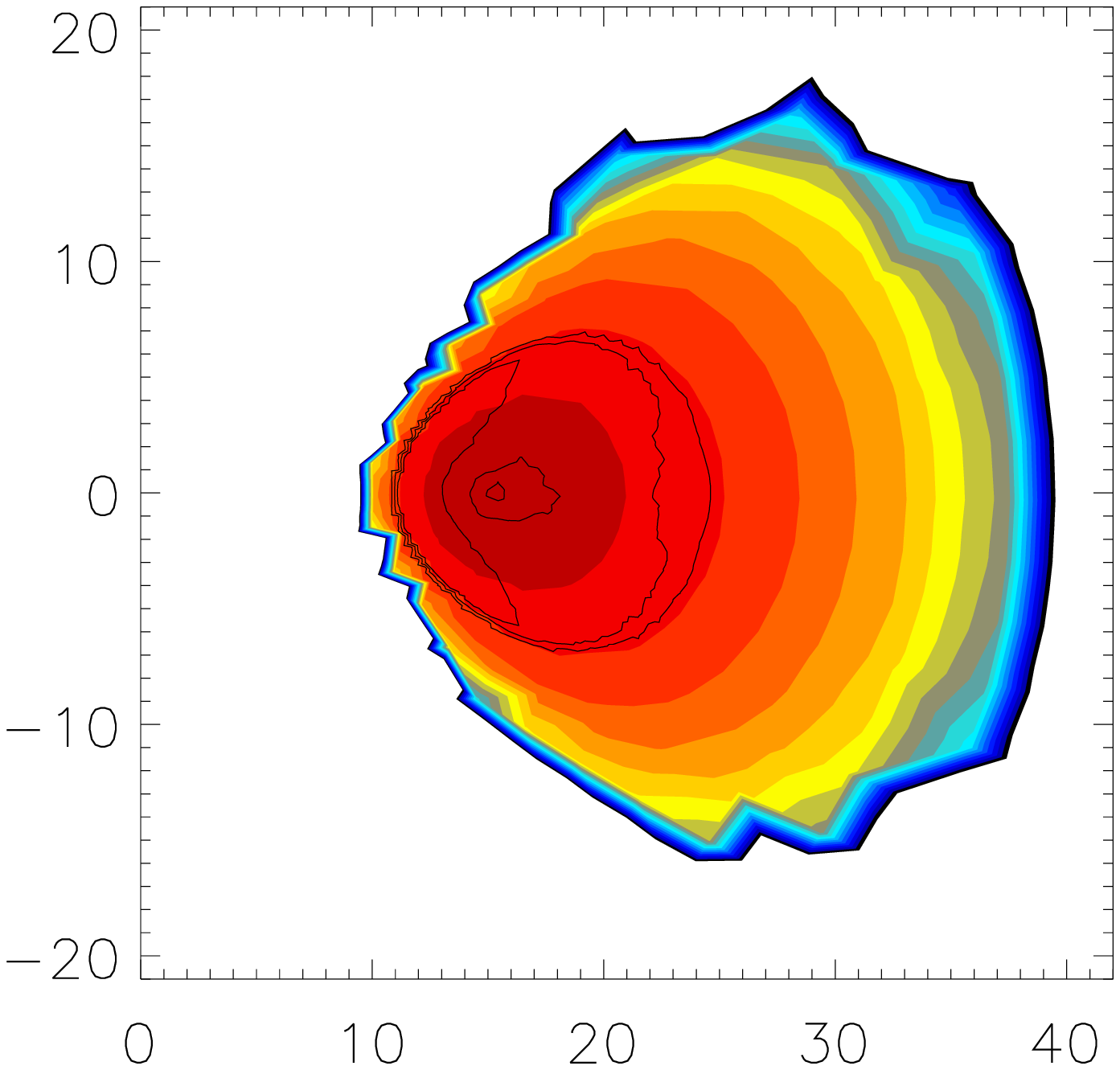}
   \caption{\label{SF_init} 
    Initial density and magnetic pressure profiles for the
    (a) SFR, (b) SF0, and (c) SFP models.  Density is plotted on a
    logarithmic scale, in units of the density maximum.  The colors
    cover four decades from red (high) to blue (low). 
    Magnetic pressure, shown as overlaid contour lines, is also 
    scaled logarithmically.}  
\end{figure}

Figure \ref{SF_init} shows the initial density profiles for these
fiducial tori, plotted on the same spatial scale to illustrate clearly
the relative size of the disks.  Overlaid on the tori are contours of
the initial magnetic pressure.  The choice of vector potential density
cut-off (eq.  \ref{vecpot}), $\rho_{cut} = 0.5\,\rho_{max}$, yields
initial poloidal loops concentrated in the immediate vicinity of the
pressure maximum.  Although this keeps the field well away from the
steep drop-off in density near the inner edge of the disk, the initial
confinement of the field to the core of the disk gives rise to a violent
transient as the field is amplified and buoyantly escapes from the core.

\begin{figure}[ht]
    \epsscale{0.335}
    \plotone{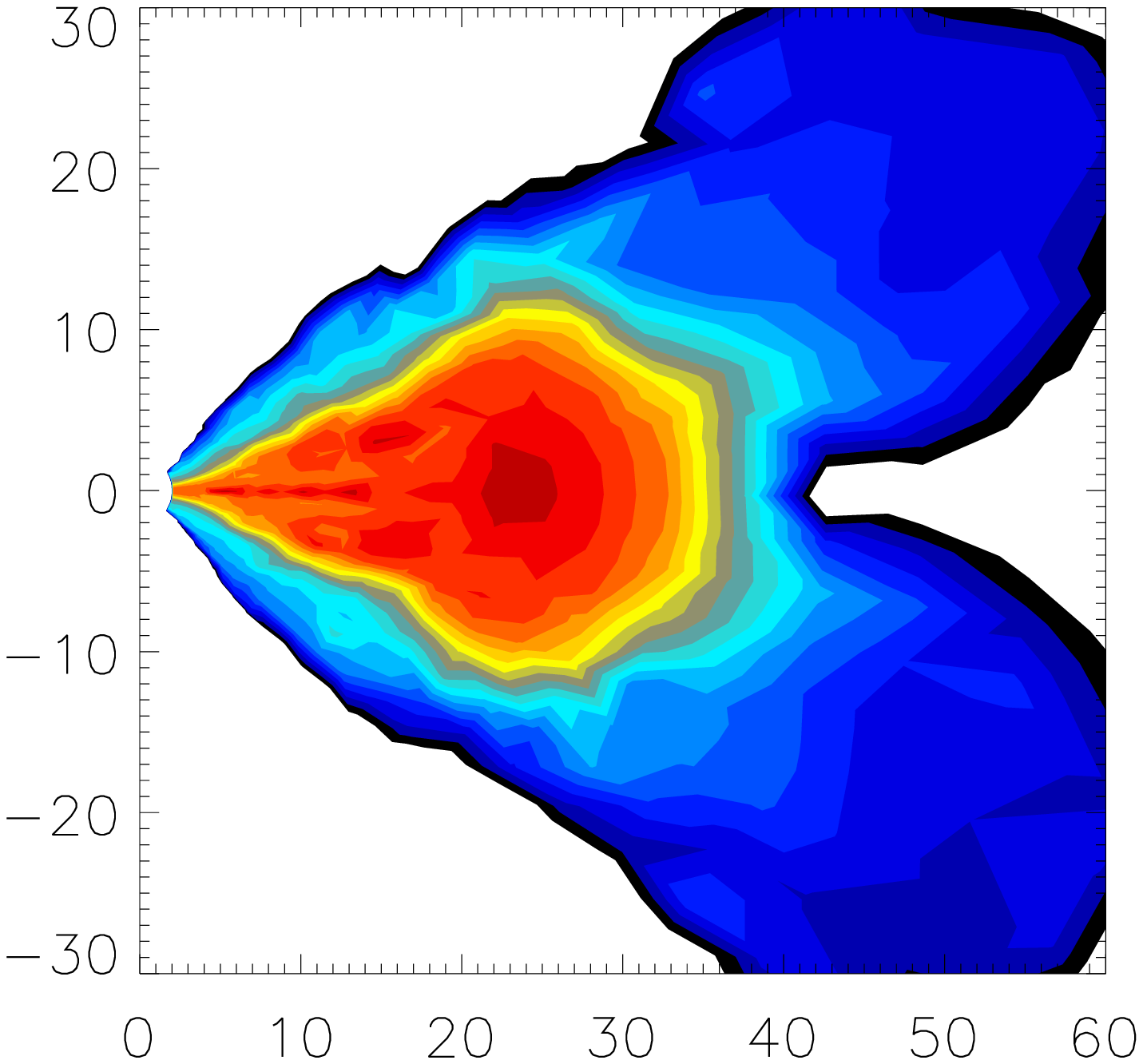}\plotone{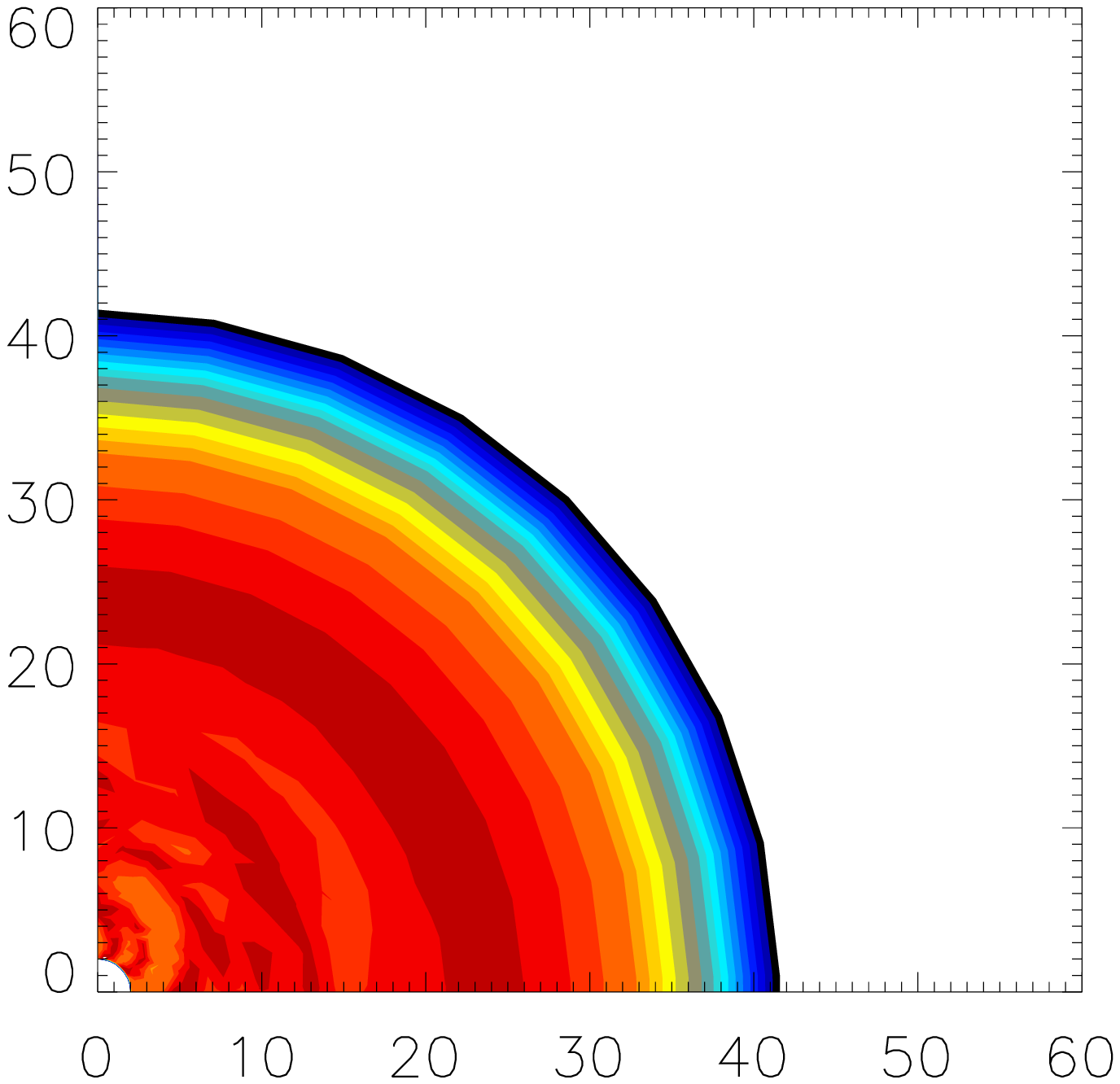}
    \plotone{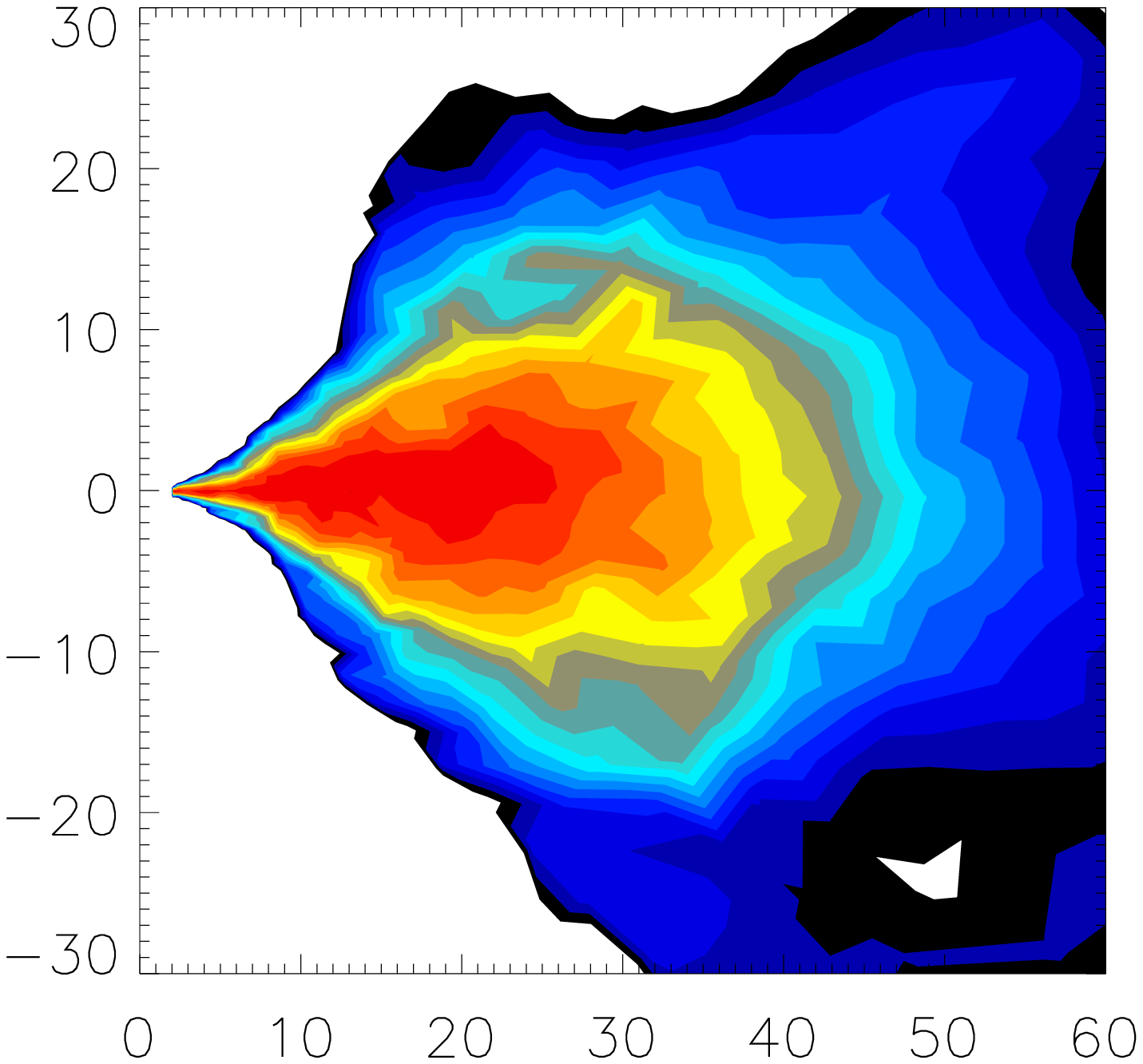}\plotone{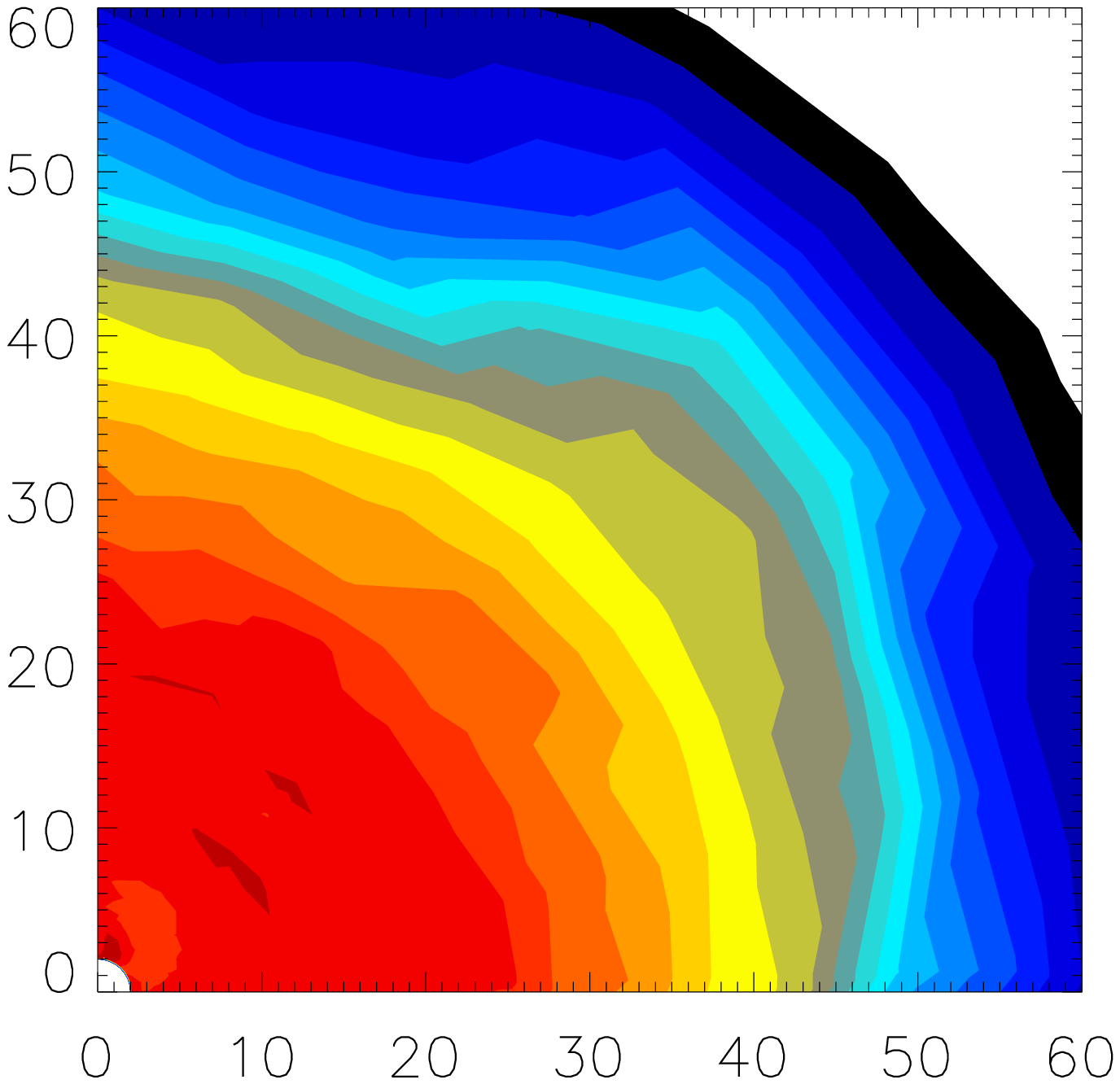}
    \plotone{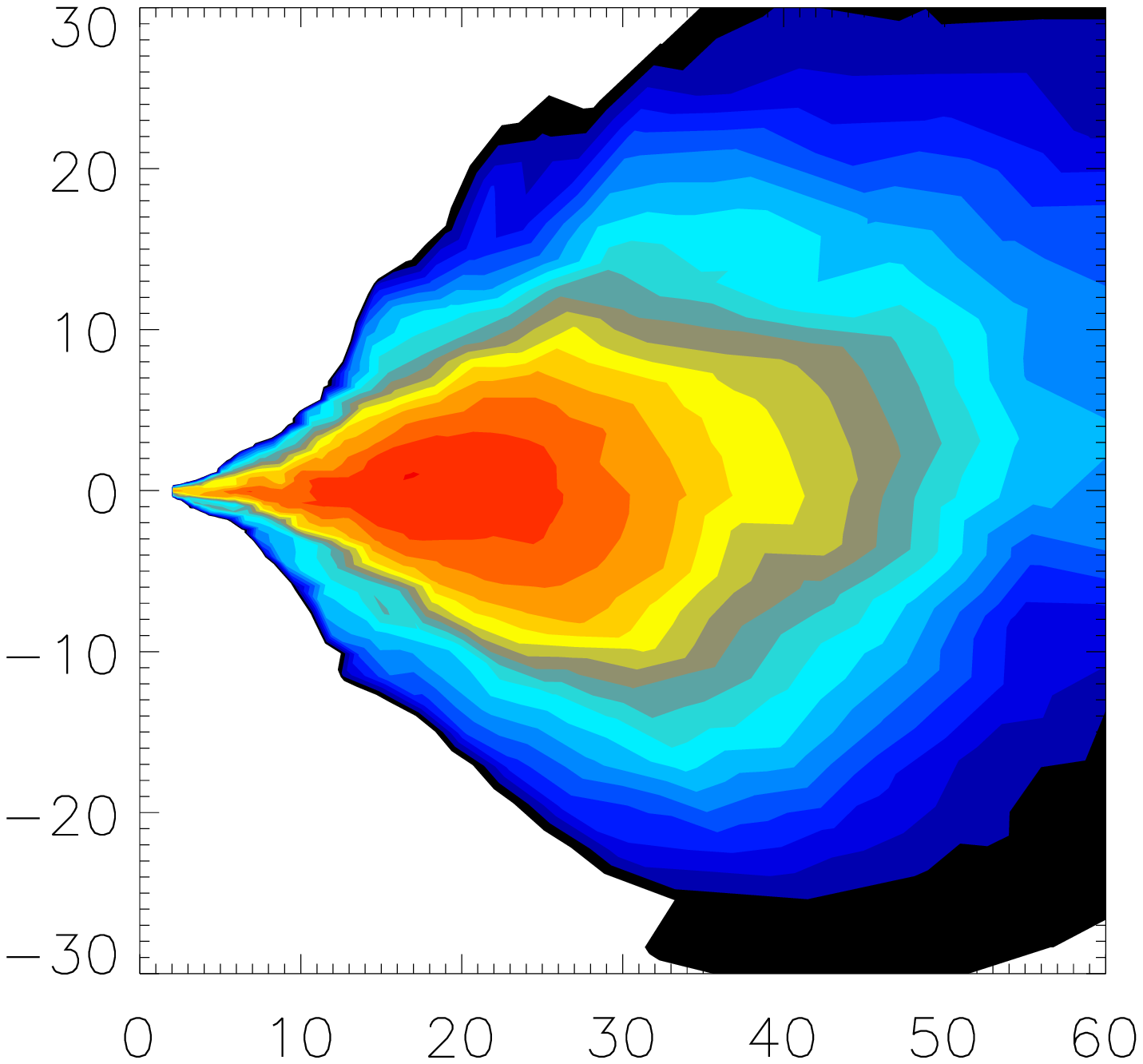}\plotone{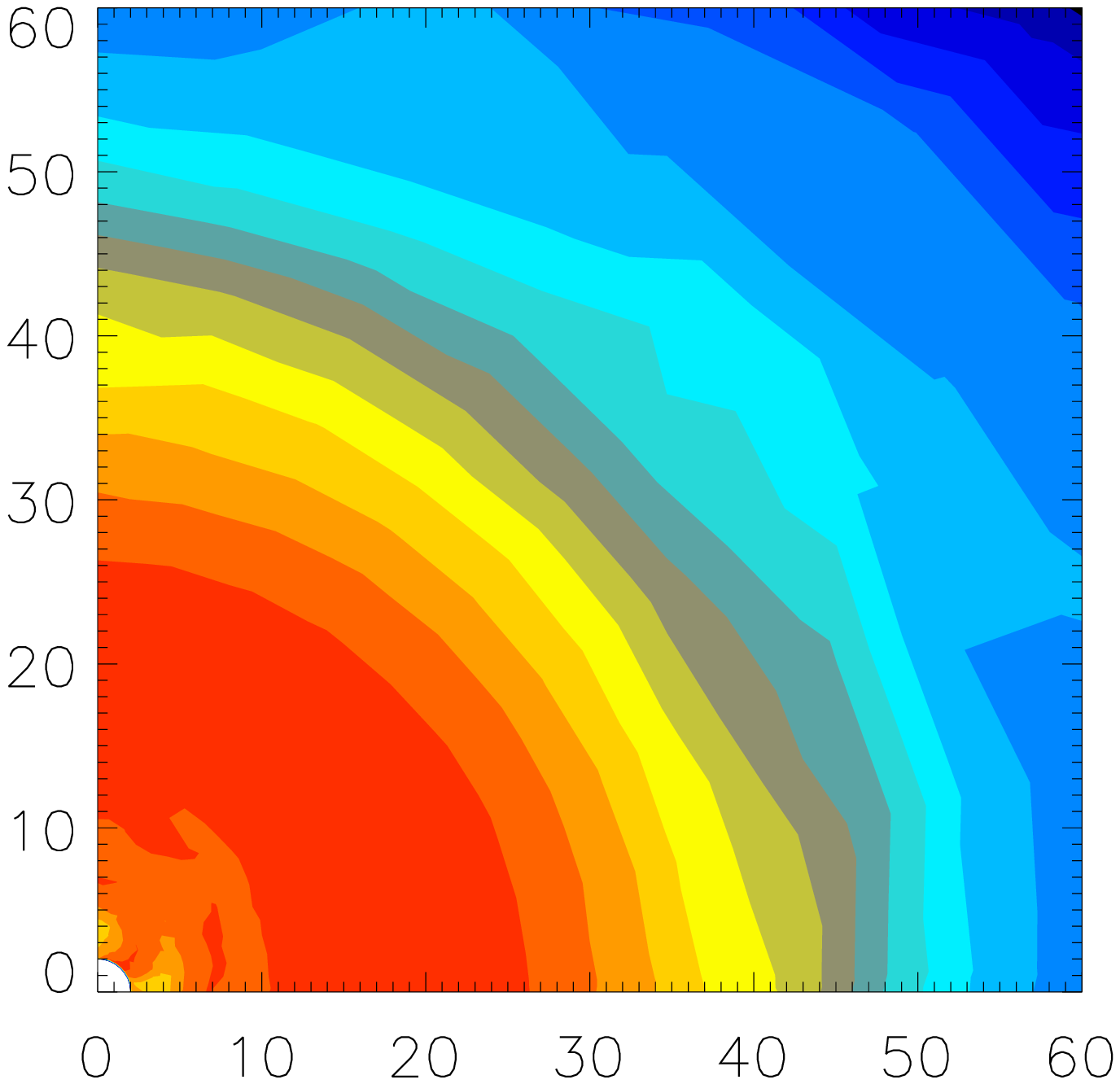}
    \caption{\label{SFMpol} 
     Plots of log density for model SFR. Panels (a), (c), and (e)
     are polar slices through the disk at $\phi=0$. Panels (b), (d), and (f)
     are equatorial slices through the disk at $\theta=\pi/2$. 
     Panels (a,b) are taken at $t=2.0$ orbits;
     (c,d) at $t=4.0$ orbits;
     (e,f) at $t=10.0$ orbits.} 
\end{figure}

\begin{figure}[ht]
    \epsscale{0.335}
    \plotone{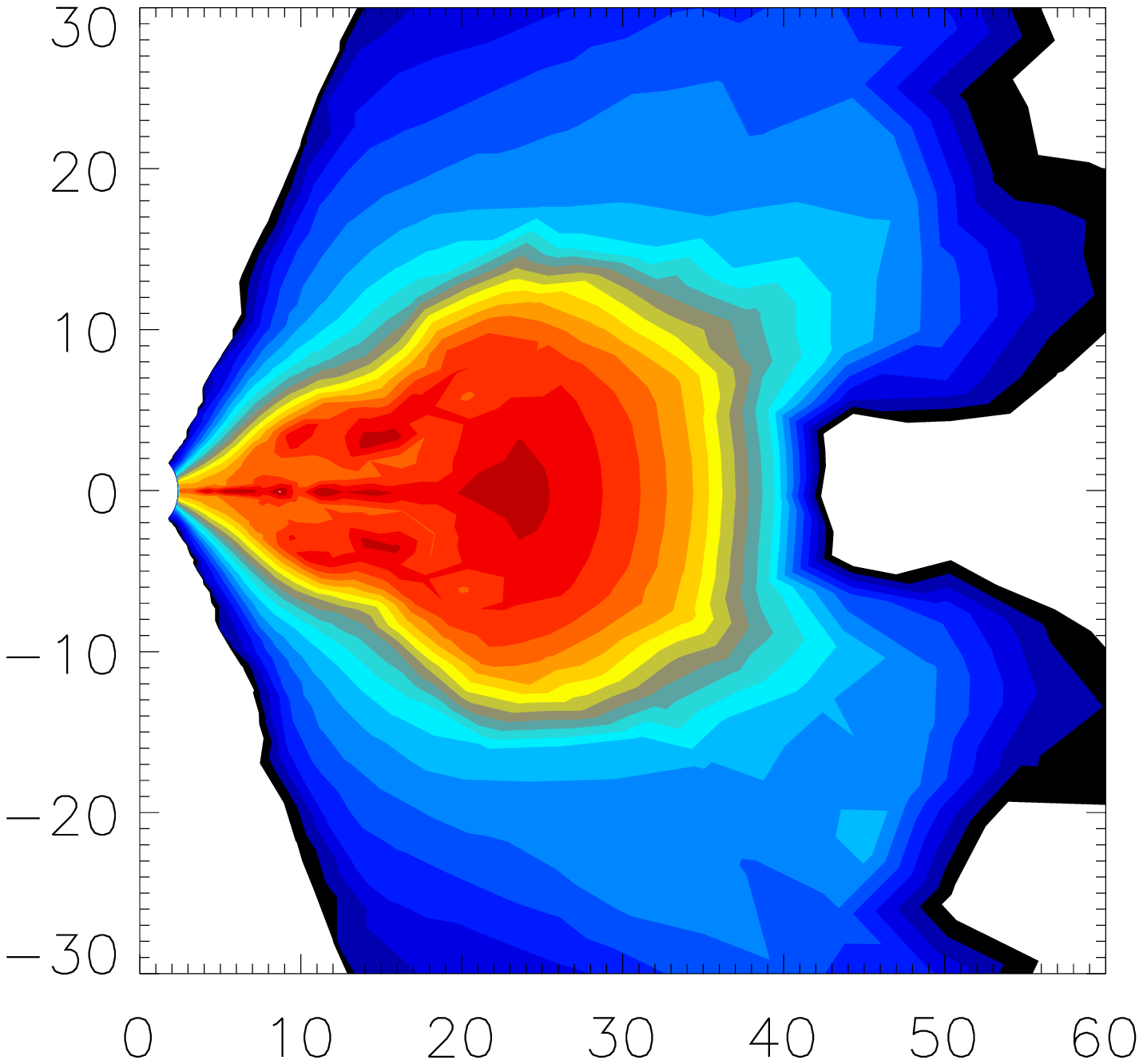}\plotone{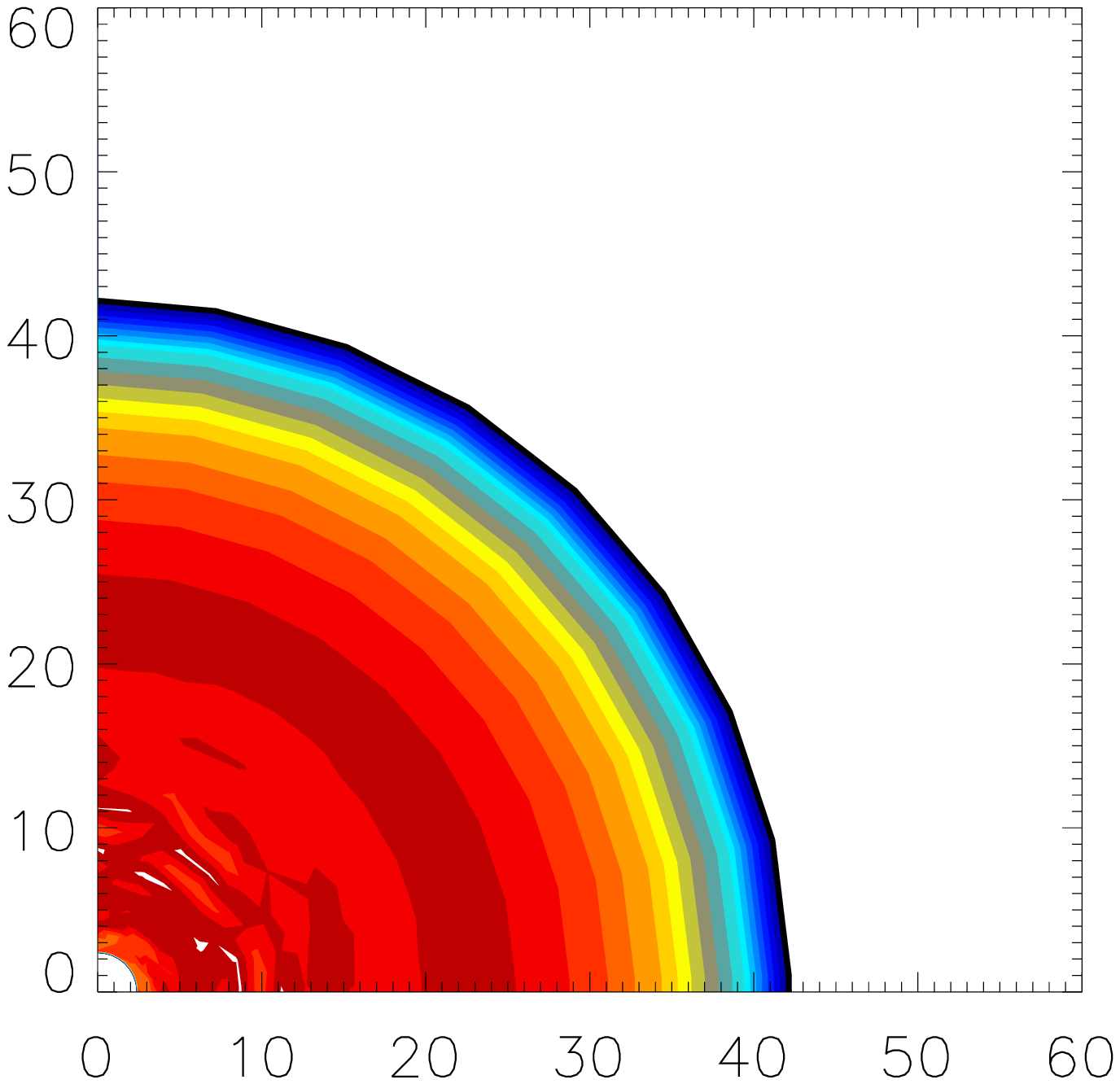}
    \plotone{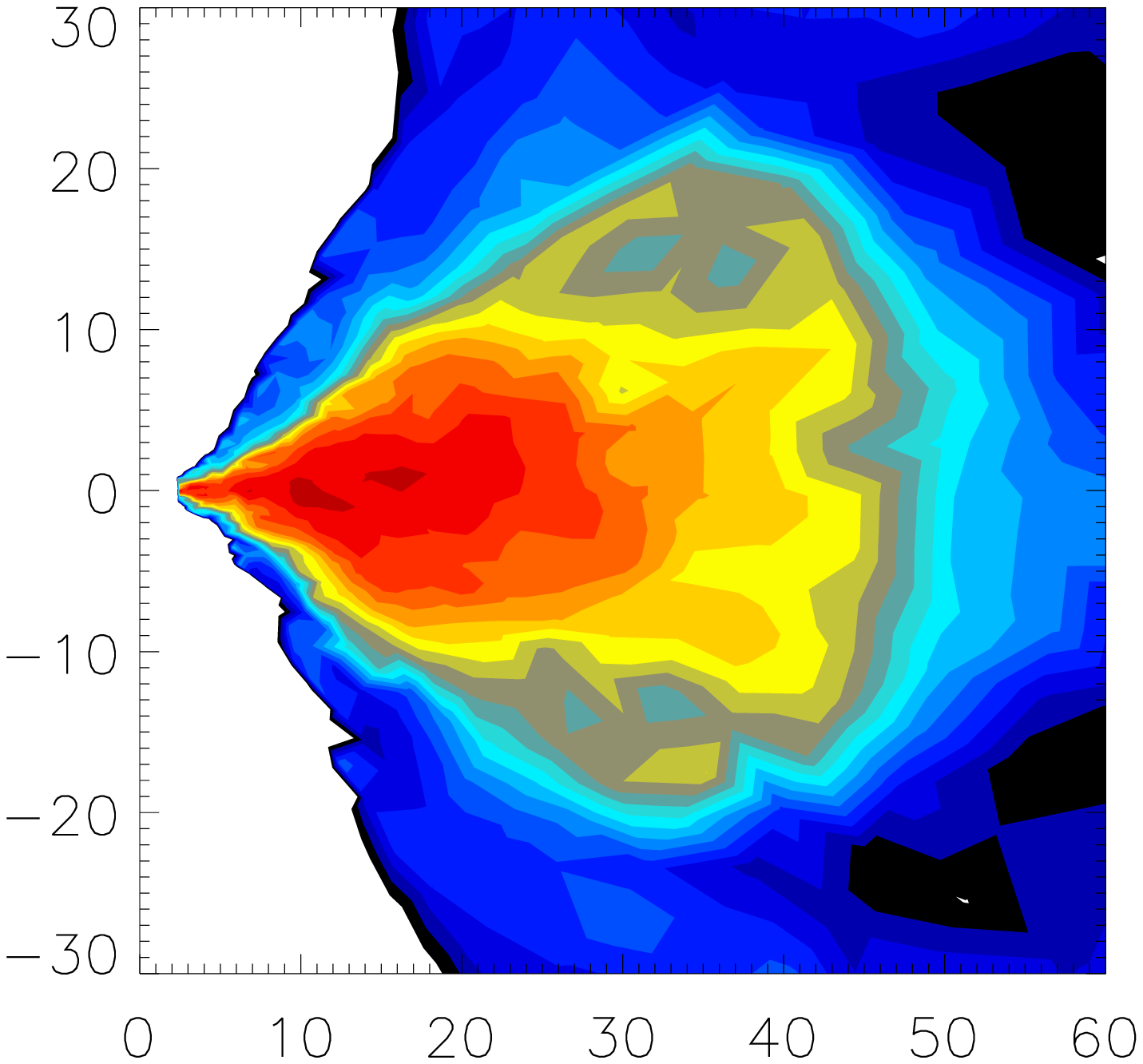}\plotone{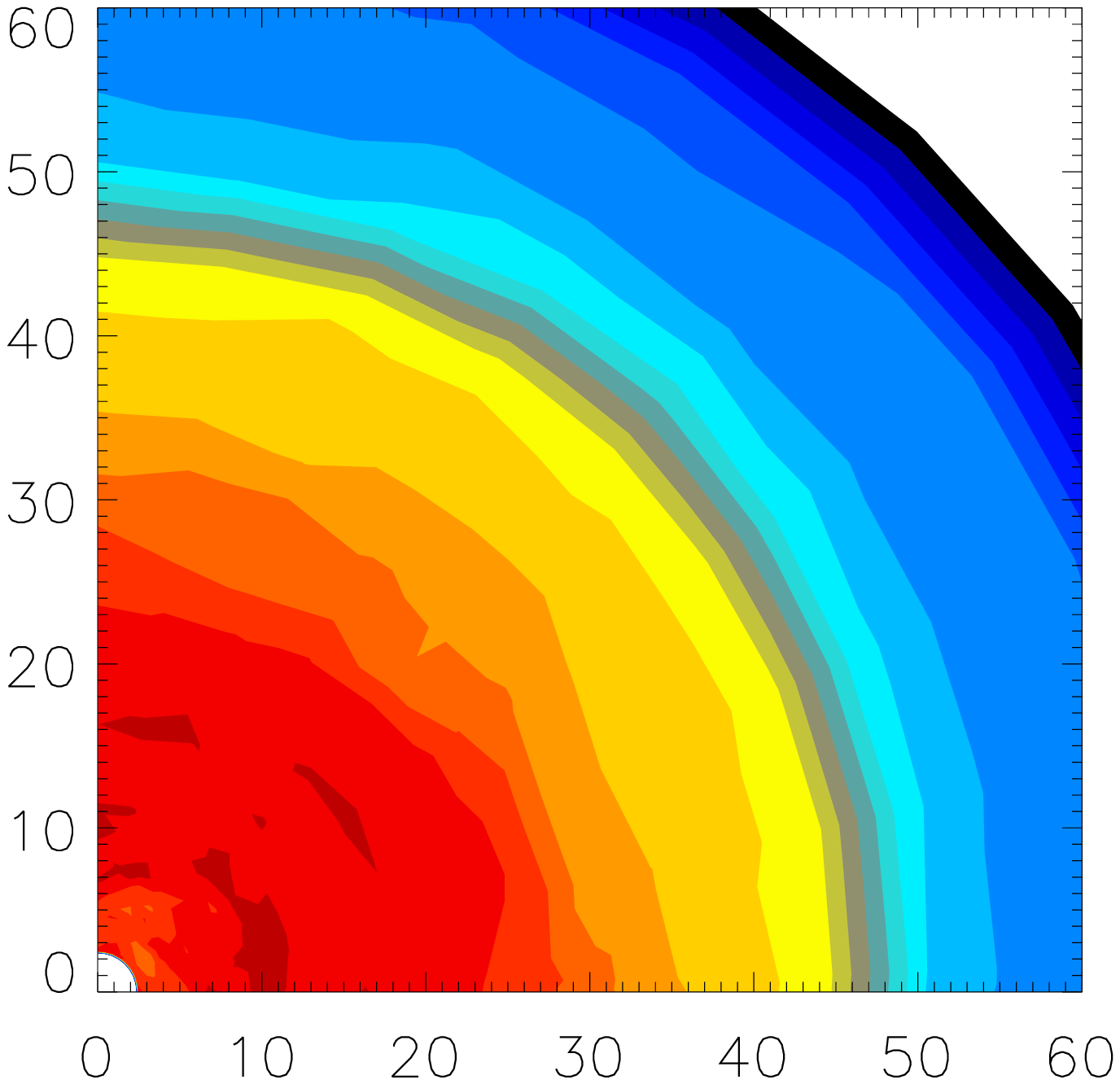}
    \plotone{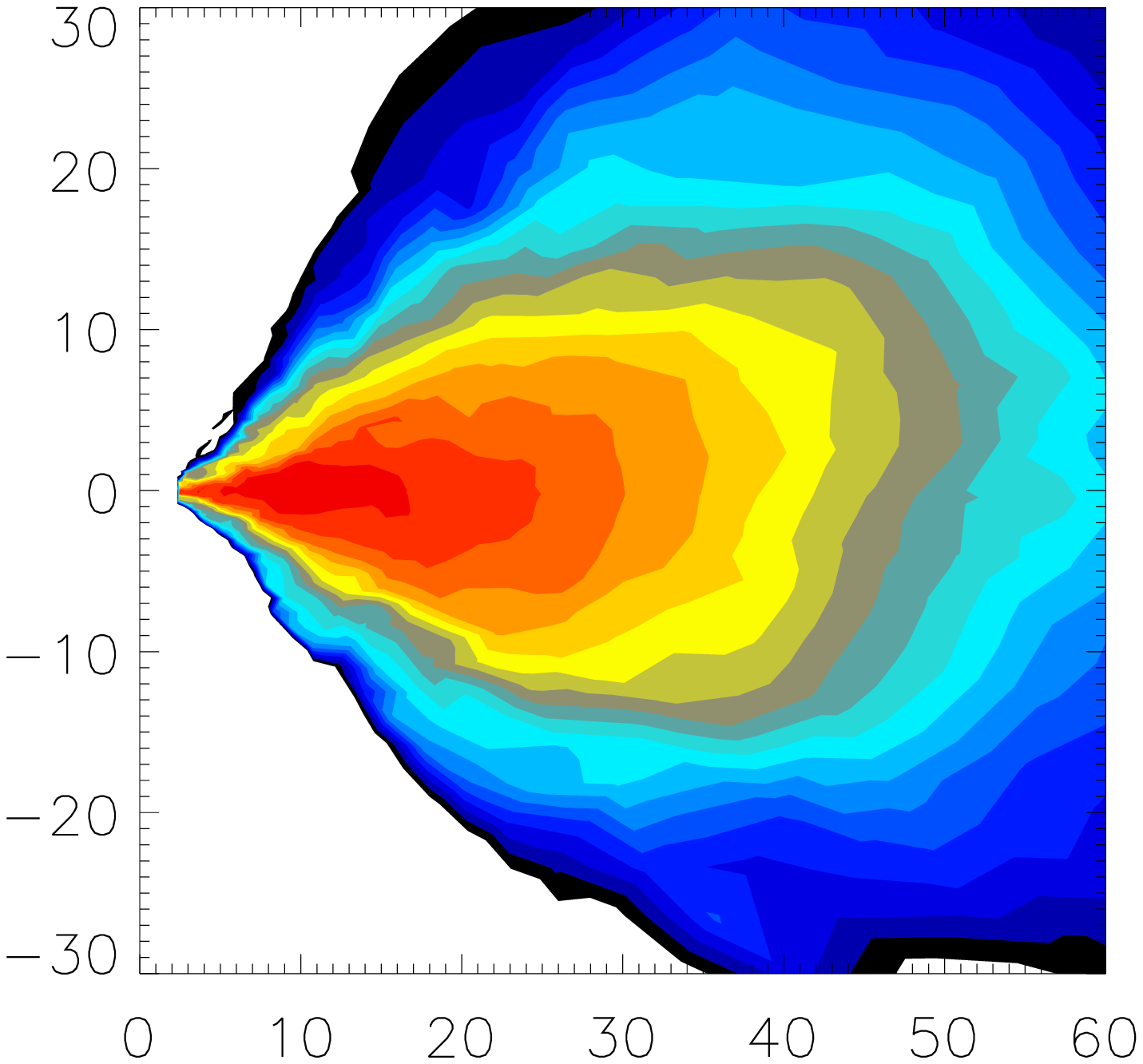}\plotone{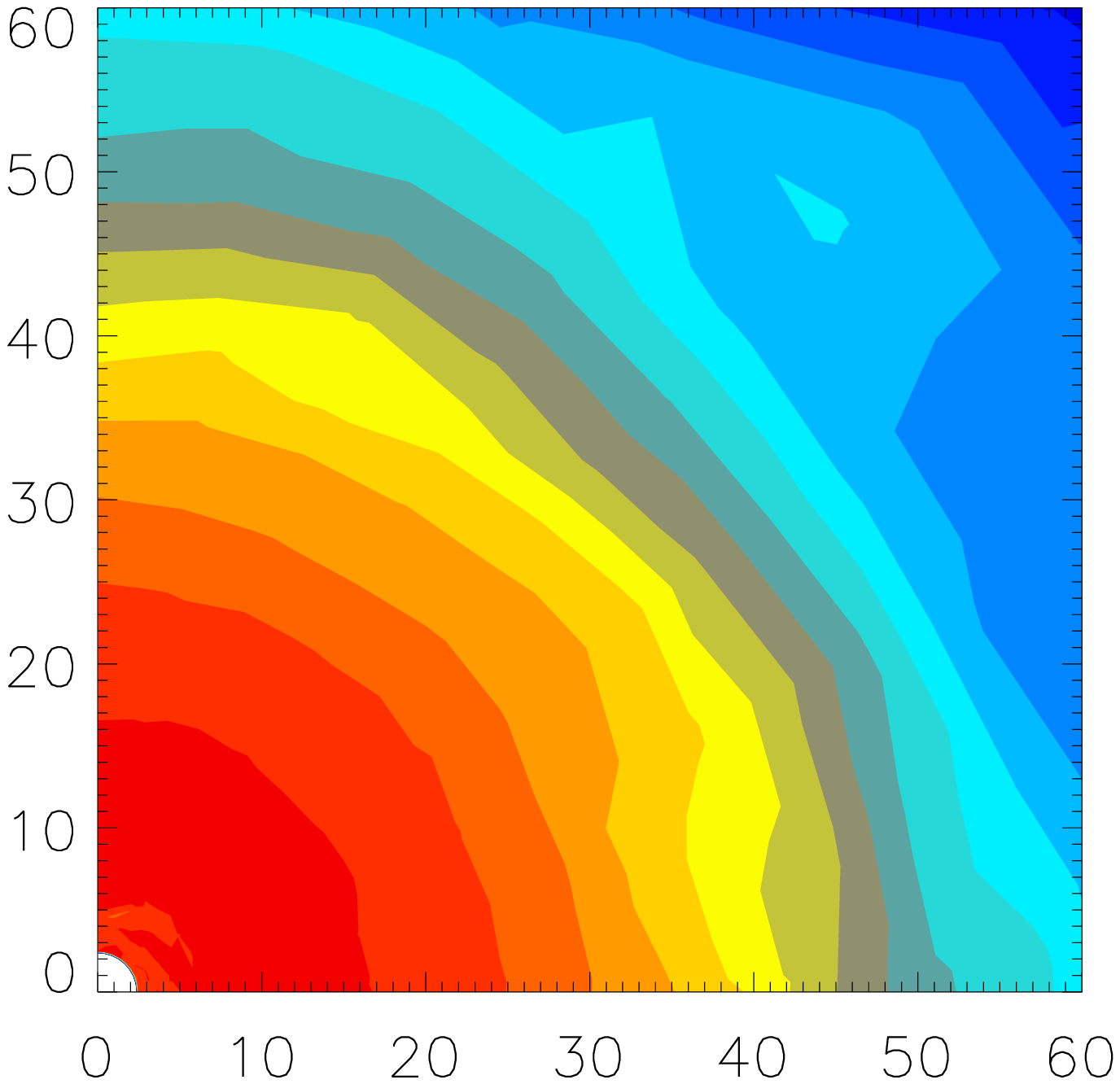}
    \caption{\label{SF0pol} 
     Plots of log density for model SF0. Panels (a), (c), and (e)
     are polar slices through the disk at $\phi=0$. Panels (b), (d), and (f)
     are equatorial slices through the disk at $\theta=\pi/2$. 
     Panels (a,b) are taken at $t=2.0$ orbits;
     (c,d) at $t=4.0$ orbits;
     (e,f) at $t=10.0$ orbits.} 
\end{figure}

\begin{figure}[ht]
    \epsscale{0.335}
    \plotone{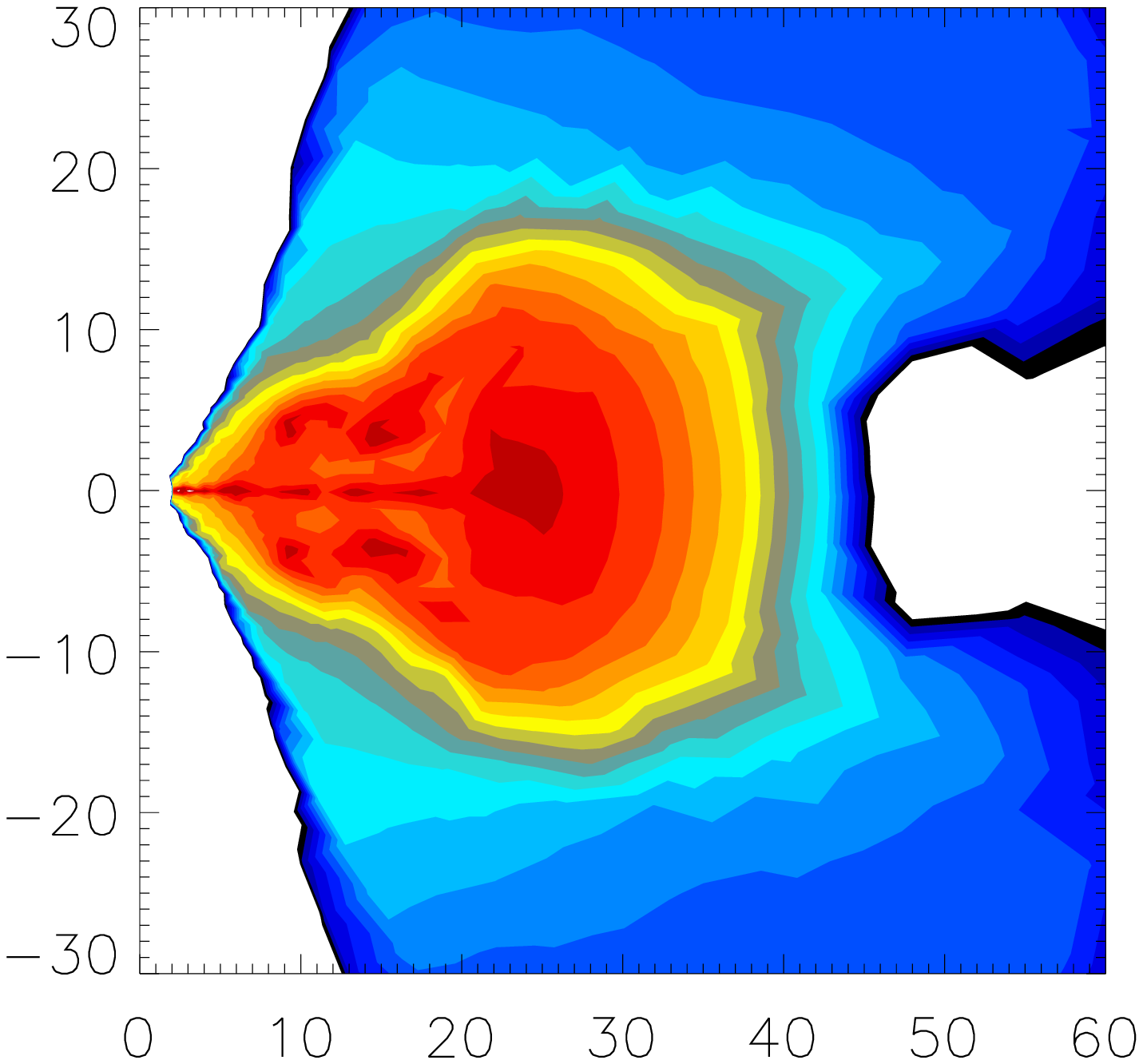}\plotone{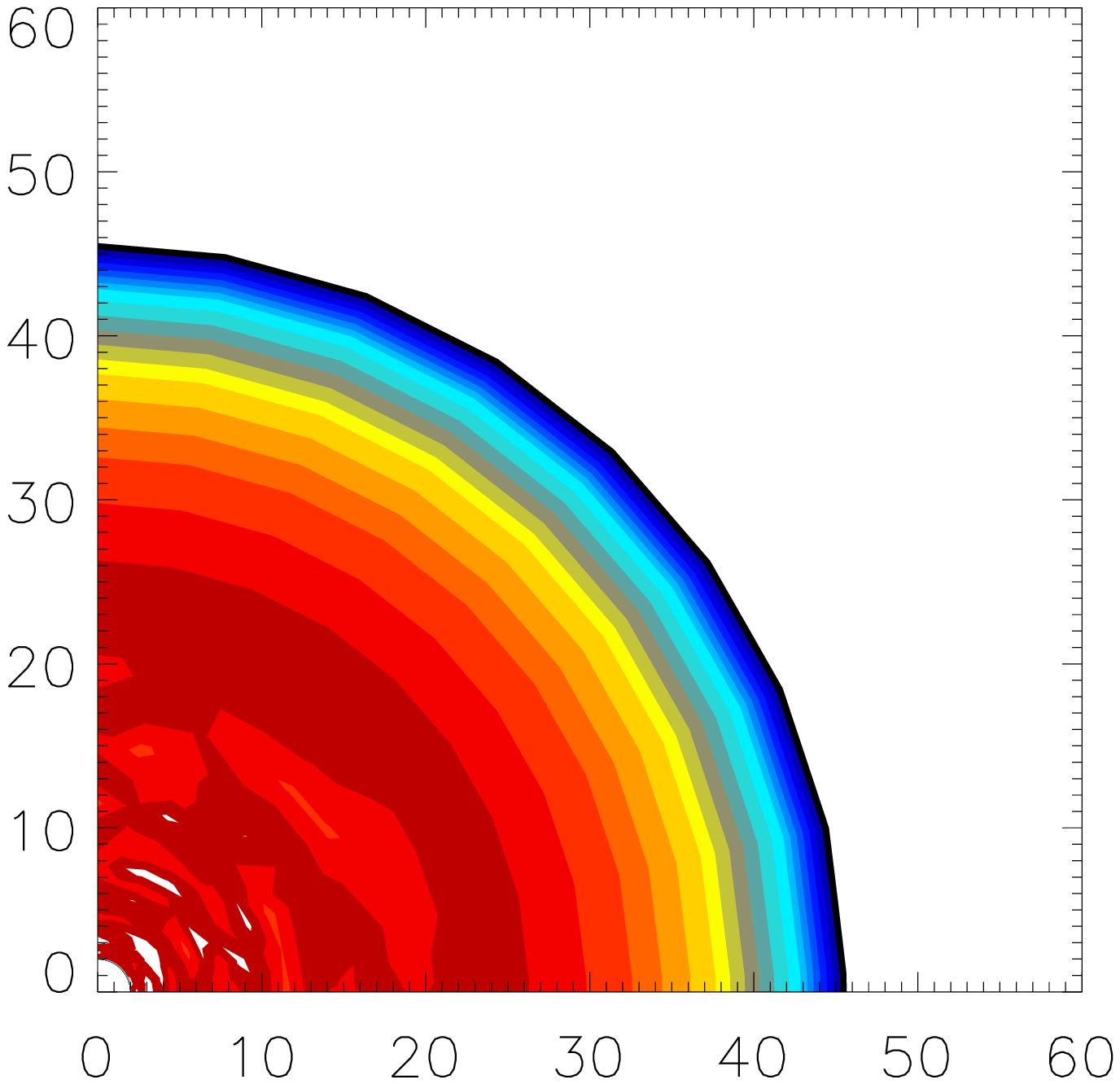}
    \plotone{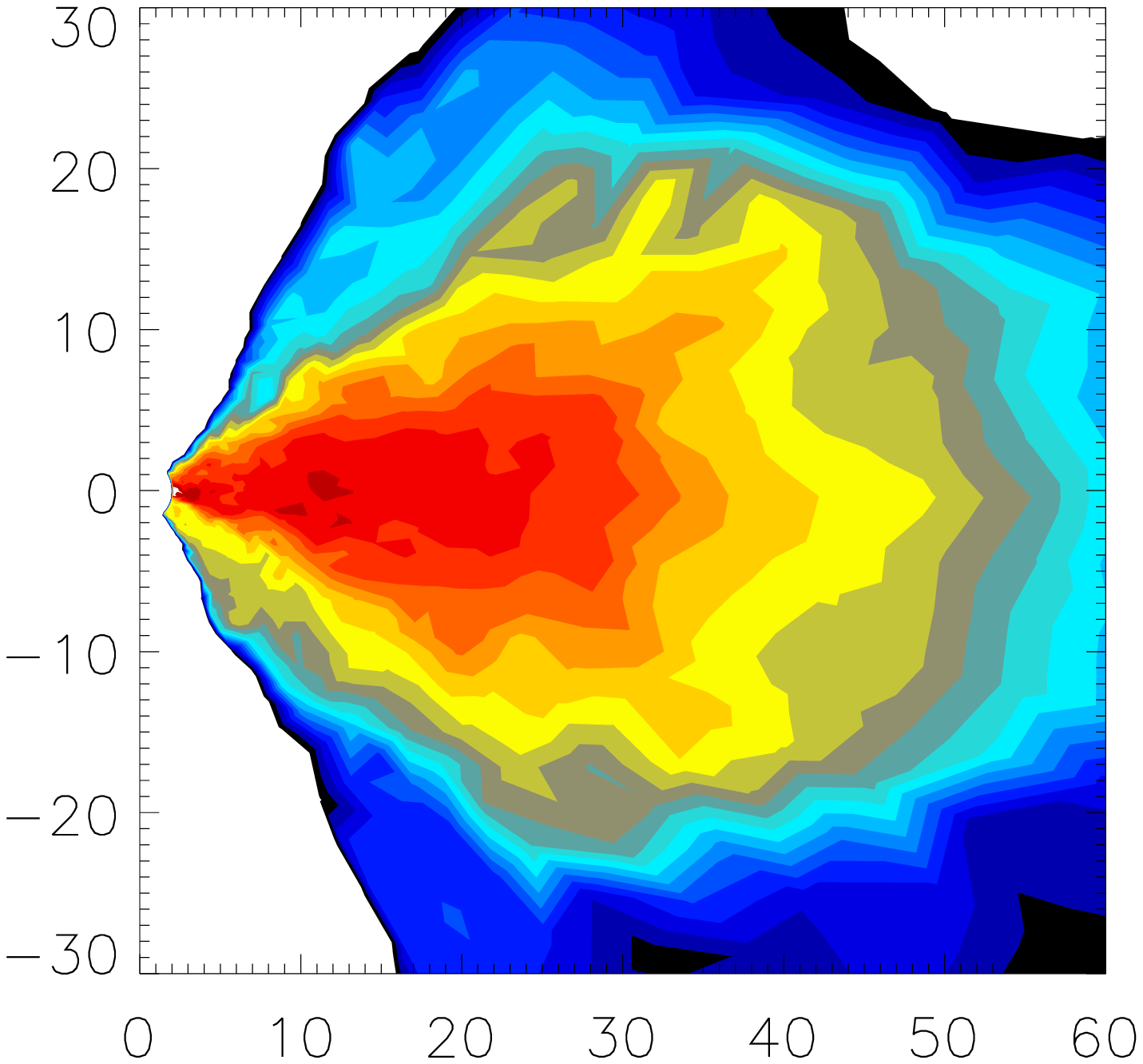}\plotone{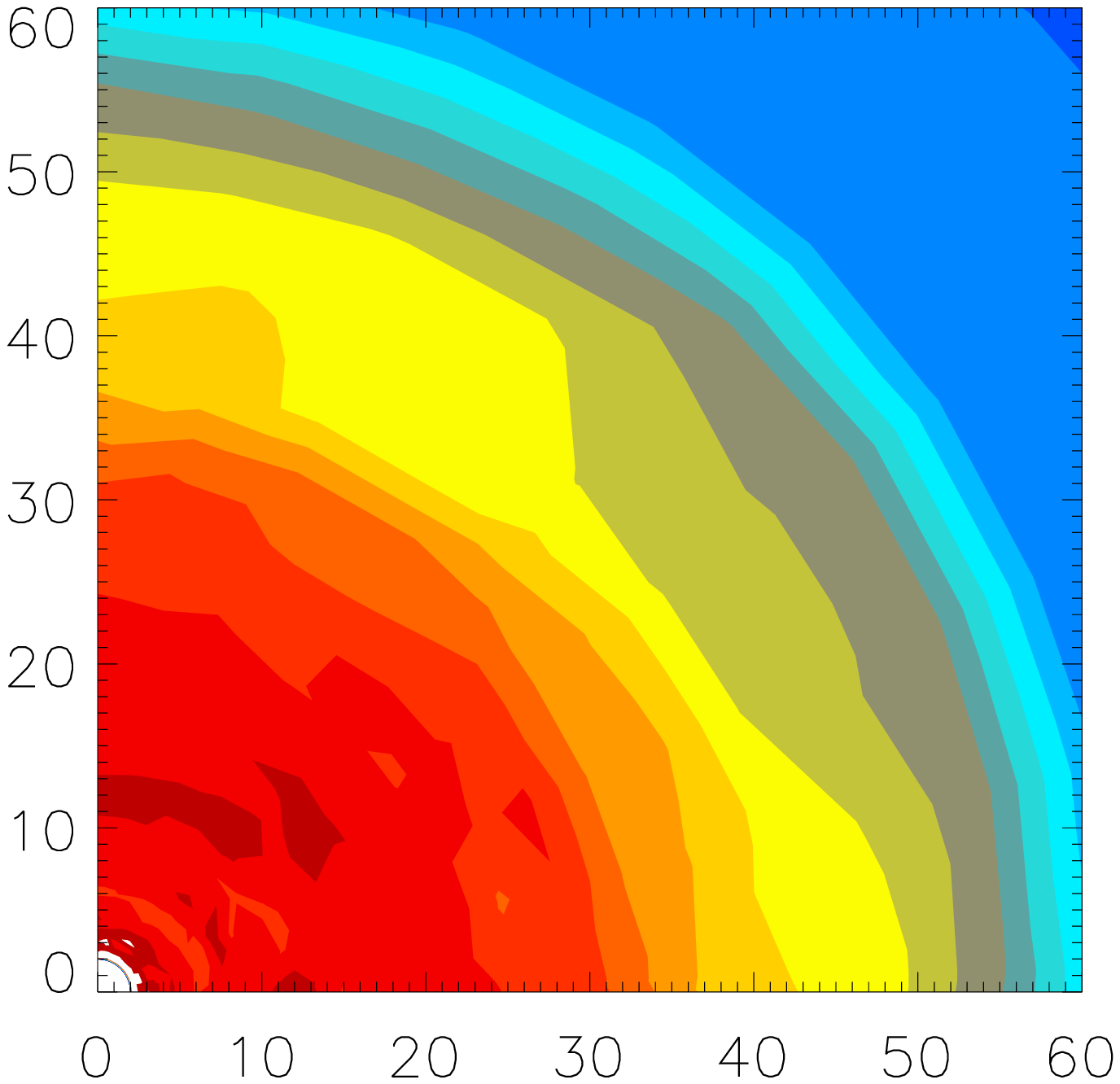}
    \plotone{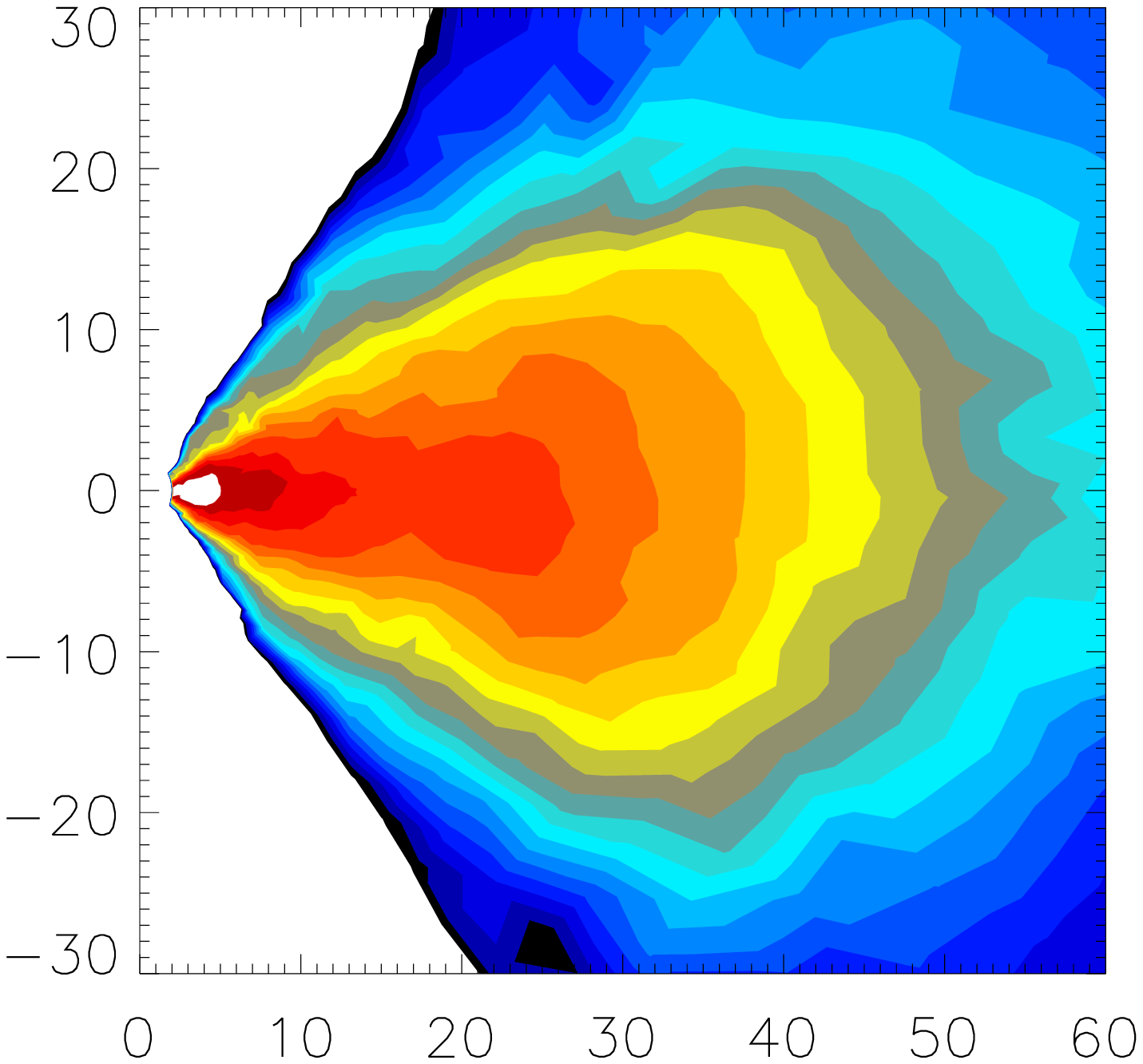}\plotone{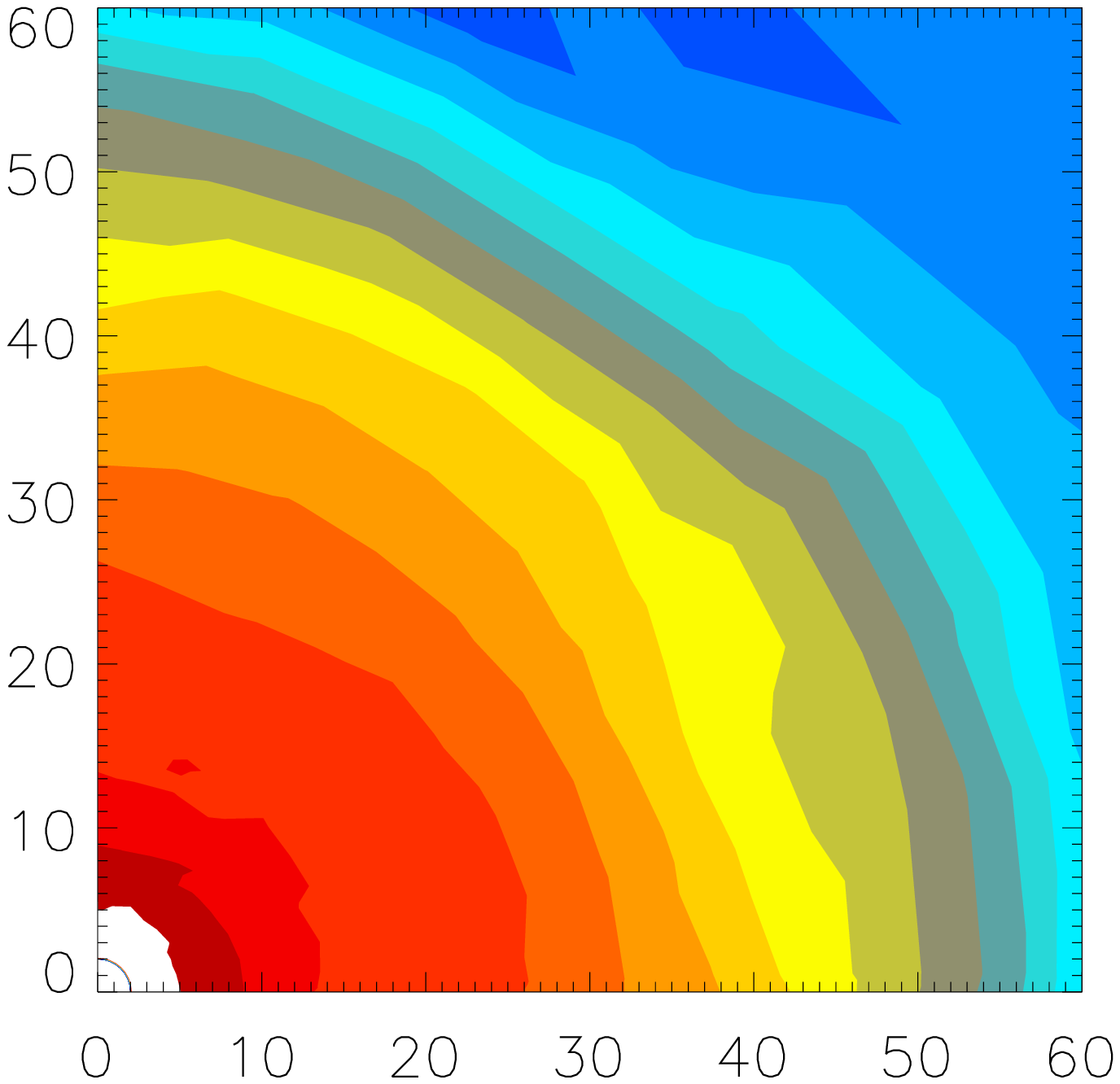}
    \caption{\label{SFPtwo} 
     Plots of log density for model SFP. Panels (a), (c), and (e)
     are polar slices through the disk at $\phi=0$. Panels (b), (d), and (f)
     are equatorial slices through the disk at $\theta=\pi/2$. 
     Panels (a,b) are taken at $t=2.0$ orbits;
     (c,d) at $t=4.0$ orbits;
     (e,f) at $t=10.0$ orbits.} 
\end{figure}

The three models, SFR, SF0, and SFP, were evolved for a time equivalent
to $10.0$ orbits at their respective pressure maxima.  Figures
\ref{SFMpol}, \ref{SF0pol}, and \ref{SFPtwo} show polar and equatorial
slices through the dataset for the density $\rho$ taken at $t=2.0$,
$4.0$, $10.0$ orbits.  During the evolution the tori pass through three
phases:  (1) the initial rapid growth of the toroidal magnetic field by
shear; (2) the initial nonlinear saturation of the poloidal field MRI;
and (3) evolution by sustained MHD turbulence.

During the first phase, the shearing of the radial field quickly
generates a significant toroidal field.  The increase in magnetic
pressure causes the inner region of the torus to expand, driving both
accretion and outflow.  The second phase, the violent nonlinear
saturation of the MRI, begins at about orbit 2 and features strong
radial motions.  Panels (a) and (b) of each figure show plots at
$t=2.0$ orbits, where this violent transient state is at its peak.  A
strong, thin current sheet is clearly visible on the equator.  A bubble
of high magnetic pressure straddles the equator, driving lower density
gas towards the black hole, while just above and below this bubble,
pincers of denser material are also driven towards the black hole.  The
equatorial slices at $t=2.0$ show a series of spirals inside $r \approx
15\, M$, that represent a well-organized flow of material towards the
hole.  In model SFR these spiral waves reverse direction near the hole
due to frame dragging.  This flow feature was also observed in the
hydrodynamic simulations of DH02.

As gas is driven inward by magnetic pressure, it also expands upward,
carrying with it significant magnetic field.  This leads to the
formation of a highly magnetized outflow which moves out above and
below the torus.  This backflow is a generic feature of geometrically
thick, nonradiating accretion flows, and has been observed in previous
pseudo-Newtonian simulations (Hawley \& Balbus 2002; Hawley \& Krolik
2001, 2002).

Through orbit 2 the flow retains a high degree of symmetry across the
equator, but when the pincers reach the black hole, the near-horizon
inflow and the inner part of the torus become turbulent, as can be seen
at $t=4.0$ orbits in panels (c) and (d).   At this stage, the outer
regions of the torus are still relatively well ordered, and two large
outbound magnetic bubbles of low density gas can be seen above and
below the equatorial plane, near the outer edge of the disk.  Shortly
after this, the entire disk becomes turbulent, and panels (e) and (f),
taken at $t=10.0$ orbits, are typical of this mature phase of the
evolution.  At this late stage, large spiral waves can also be seen
moving through the outer regions of the disks.

\clearpage

\clearpage

\begin{figure}[ht]
    \epsscale{0.4}
    \plotone{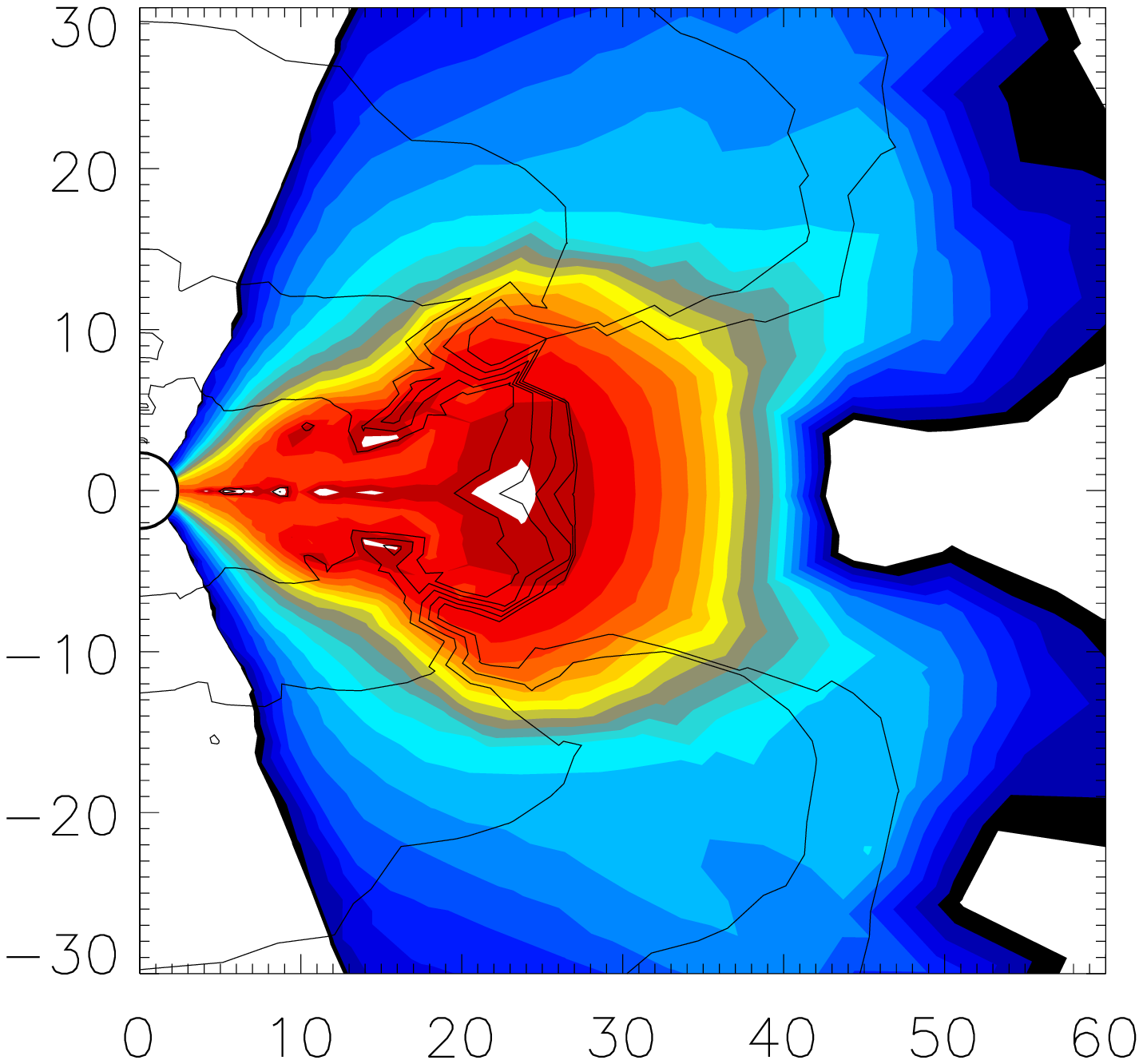}
    \plotone{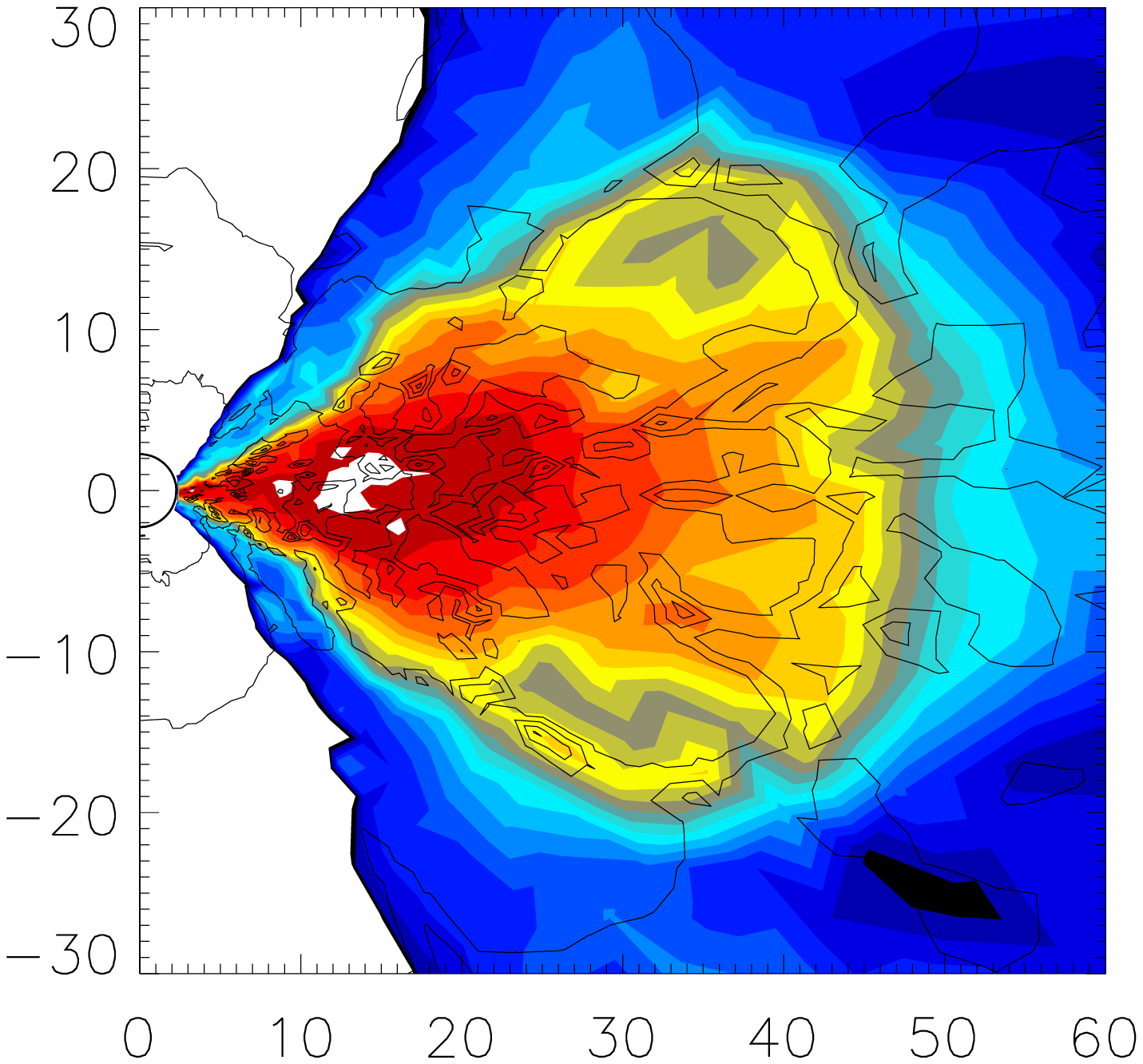}
    \caption{ \label{SF0bubbles} Magnetic bubbles in Model SF0.
     Depicted are color plots of log density at (a) $t=2.0$
     and (b) $t=4.07$ orbits, overlaid with contours of 
     magnetic pressure. The low density bubbles correspond to local 
     maxima in magnetic pressure.} 
\end{figure}

The magnetic nature of the low density bubbles at $t=2.0$ and $4.0$
orbits seen in figures \ref{SFMpol}--\ref{SFPtwo} is made clear by
Figure \ref{SF0bubbles}, which shows plots of the density in the SF0
model at $t=2.0$ orbits.  Figure \ref{SF0bubbles}a is at $t=2.0$
orbits and \ref{SF0bubbles}b corresponds to 4.07 orbits, a time when
the bubbles are especially prominent.  The figure shows a color plot of
density with logarithmic contours overlaid with contours of constant
magnetic pressure (dark lines).  To highlight the edge of the bubbles,
only contours of low magnetic pressure are shown (these range from
$10^{-4}$ to $10^{-3}$ of the maximum magnetic pressure on the grid).
The regions inside the density-minimum bubbles correspond to local
magnetic pressure maxima.

The general evolutionary histories of the three models are similar:
growth of a strong current sheet along the equator in the first orbit,
initial saturation of the poloidal field MRI between the second and
fourth orbit, and sustained turbulence thereafter.  The three models
also have important differences.  The most apparent difference lies in
the structure of the inflow near the horizon.  At one extreme,
retrograde model SFR has a slender inflow with decreasing pressure and
density, whereas prograde model SFP develops a new pressure maximum
region inside $r \approx 5\,M$.  We refer to this structure as the
mini-torus for its appearance when viewed in a polar slice.  Model SF0
is an intermediate case.  These differences can be understood as
arising from the relative distances between the inner edge of the
initial torus and the marginally stable orbit.  This distance is the
greatest for SFR (see Table \ref{params}) and the least for SFP.

\begin{figure}[ht] 
\epsscale{0.3} 
\plotone{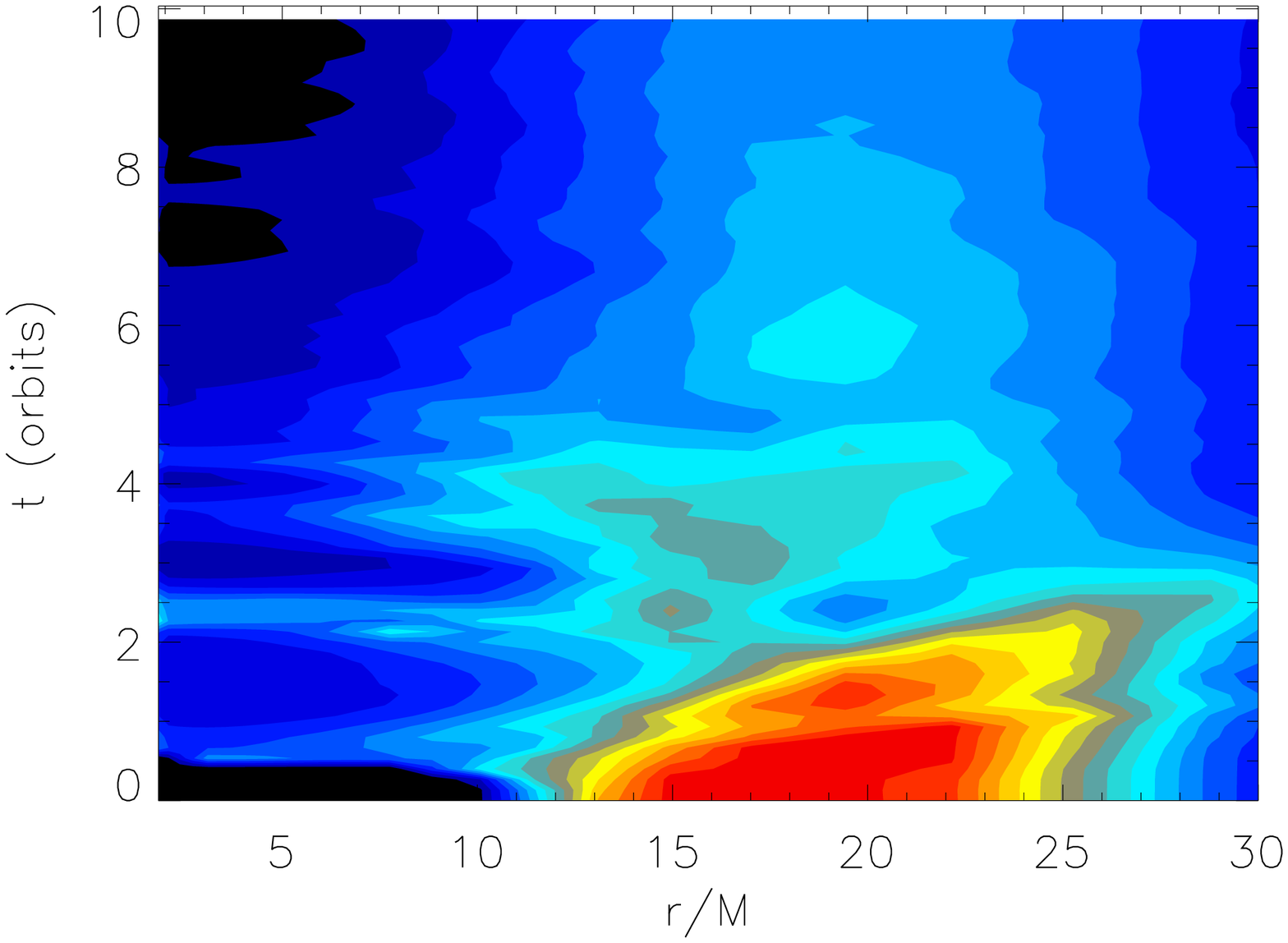}
\plotone{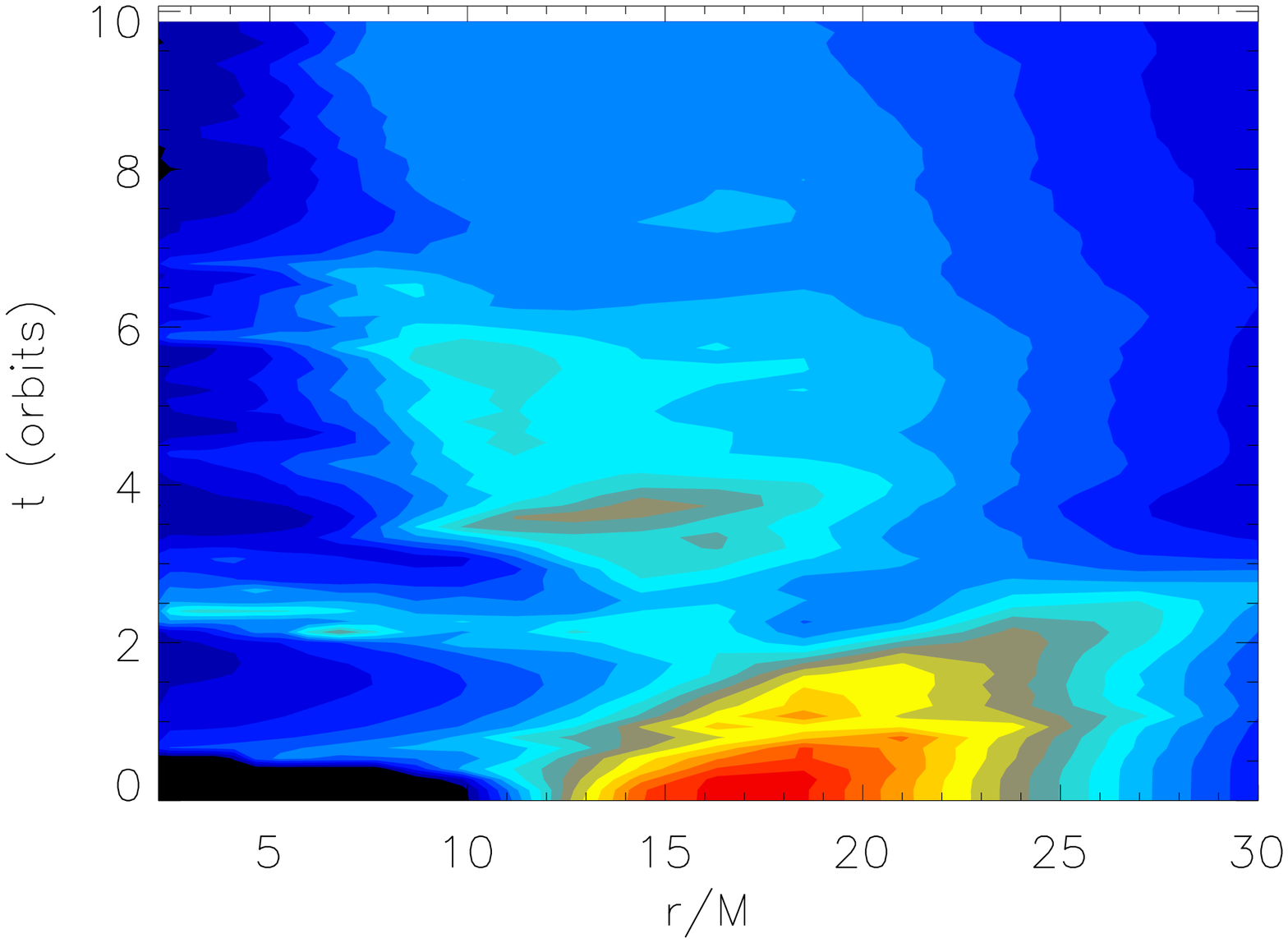}
\plotone{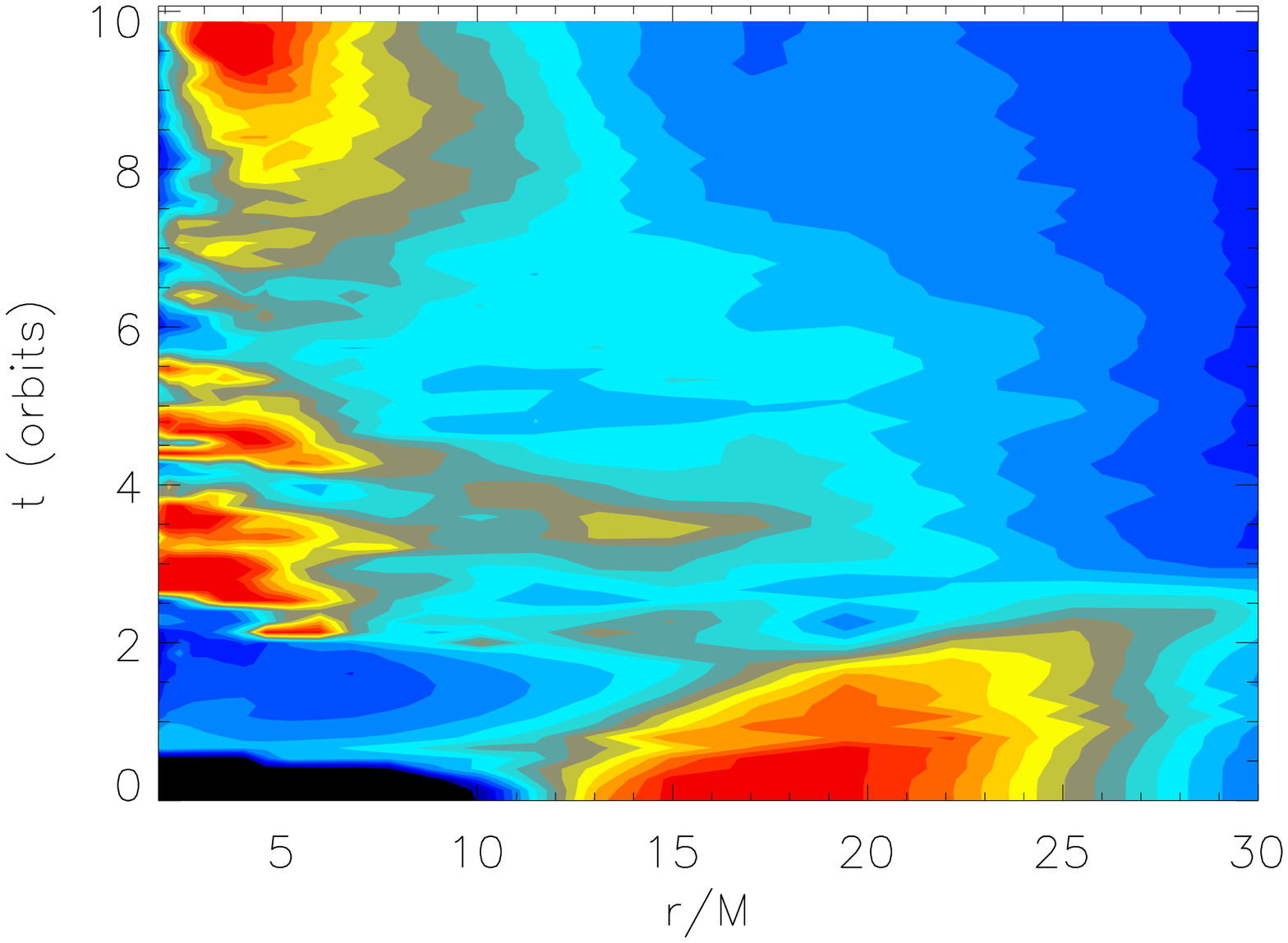}
\caption{\label{rhoav}
     Space-time diagrams of shell-averaged density, $\langle\rho\rangle$, 
for models (a) SFR, (b)
SF0, and (c) SFP. Scale is linear, and each plot is normalized to the
maximum density in the initial torus.} 
\end{figure}

A more detailed history of the evolution of the tori can be gained from
Figure \ref{rhoav}, a space-time diagram $\langle \rho\rangle (r,t)$ for
models SFR, SF0, and SFP.  The first orbit in each model shows the
result of increasing magnetic pressure:  gas is driven in toward the
hole and the density in the heart of the torus decreases.  Outgoing
waves result from backflow driven from the inner region of the torus by
this increased pressure.  After orbit 2 there are strong fluctuations
within the torus, and a series of time-variable accretion events.   In
models SF0 and SFP there is a notable shift of the denser edge of the
torus toward their appropriate marginally-stable orbits, and hence
toward the black hole.  In model SFR the denser material remains
outside of $r=10M$ except for occasional accretion streams.  After
orbit 6 the evolution is less violent.  The formation of the mini-torus
in model SFP, which appears after this point in time as a high-density
region inside of $r \approx 5\,M$, is especially obvious.

The spacetime diagram also reveals many outgoing waves.  The first of
these,  which appear in the linear MRI growth stage (between 0 and 3
orbits) originate as the backflow from the accreting gas.  From the
diagram we can measure the speed of propagation of this flow at $v
\approx 0.07\,c$.  A prominent tilted strip, visible in each plot
between $t \approx 1$ and $3$ orbits, is associated with outward moving
material propagating from inner edge of the inflow along the surface of
the disk; this feature is especially prominent in animations of the
data.  A series of weaker outgoing waves are visible at later times.
These are often internal spiral pressure waves rather than bulk
outflows.  These waves are also visible in animations of the data, and
originate deep within the turbulent accretion flow.  In model SFP, for
example, strong waves originate within the mini-torus.

\begin{figure}[htb]
    \epsscale{0.4}
    \plotone{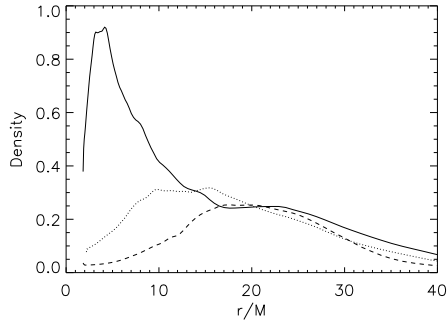}
    \caption{\label{rhoavseq}  
     Plot of $\langle\rho \rangle$ as a function of radius for models 
     SFR (dashed line), SF0 (dotted line)
     and SFP (solid line) at $t=10$ orbits.  The density is
     normalized to the maximum value in the initial torus.} 
\end{figure}

A quantitative view of the density distribution is provided in Figure
\ref{rhoavseq}, which shows $\langle \rho \rangle$ for each disk model
at $t=10$ orbits.  In each case the density begins to decline outside
of the location of the marginally stable orbit.   In model SFP $r_{ms}=
2.32M$, and the density increases roughly as $1/r$ inside the initial
torus's inner boundary.  The region of peak density, lying between
$r=3M$ and $5M$, is the region referred to as the mini-torus.

\begin{figure}[ht]
    \epsscale{0.4}
    \plotone{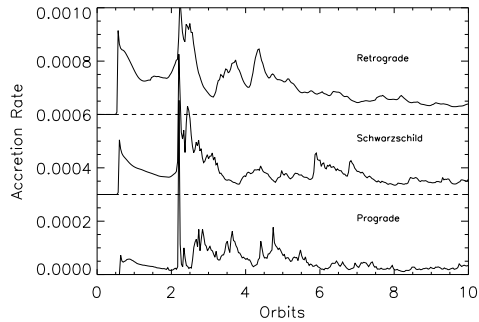}
    \caption{\label{Mflux} 
     Accretion rate, $\dot M$, at the inner radial boundary 
    ($r_{min}$).  The accretion rate is normalized by the initial
    torus mass.   The value for SFR (Retrograde) is offset by 0.0006,
    and SF0 (Schwarzschild) is offset by 0.0003 for clarity.}
\end{figure}

Figure \ref{Mflux} shows the accretion rate through the innermost
radial zones, $\dot M = -\langle \rho U^r\rangle (r_{min},t)$, normalized
to the initial torus mass for each of the models.  The violent
transient that results from the onset of the strong nonlinear phase of
the MRI is clearly visible as a sharp spike at $t=2.2$ orbits in all
three curves.  This spike corresponds to a strong radial streaming flow
of short duration.  This type of flow is typical for the early
nonlinear stage of the MRI operating on a background poloidal field.
Strong fluctuations continue for two orbits, but by the fifth orbit all
models settle into a steadier overall accretion rate with fluctuations
about the mean on short time scales.  This figure shows that, as a
function of initial torus mass, model SFR has the greatest accretion
rate, and model SFP the least.  Accretion is easier to accomplish in
model SFR because the initial torus inner boundary is just outside the
location of the marginally stable orbit.  In SFP, on the other hand,
the disk must work its way down almost to $r=2M$ before reaching the
marginally stable orbit.  This is a large enough distance that the flow
must establish a new, extended turbulent disk structure.  The
mini-torus is the innermost region of that new disk.  Comparing the
three accretion rate curves also reveals that model SFR has less
variability on short timescales compared to the other models.
Variability in the accretion rate is determined by the rate at which
matter is fed in from the turbulent disk.  Fluctuations in $\dot M$,
therefore, originate mainly in the turbulent disk, outside of $r_{ms}$,
and the resulting time-dependence of $\dot M$ reflects the turbulent
frequencies of the inner region of the disk.  These frequencies are
naturally lower in the SFR model.  The accretion rates in these models
were sampled only 30 times per orbit, however, which makes it difficult
to quantify the variability, or to distinguish the variability seen in
SF0 from that of SFP.

Table \ref{masstotals} accounts for the total mass in each model.  The
values are normalized to the initial mass of the SF0 torus.  The total
initial mass, $M_0$, and the mass at 10 orbits, $M_{\rm{final}}$, are
obtained by volume integration.  The mass that enters the black hole,
$\Delta M_{\rm{in}}$, and the mass that leaves through the outer
boundary, $\Delta M_{\rm{out}}$, are obtained by integrating the mass
flux over time at the respective radial boundaries; the integral is
approximate since the flux is sampled only 30 times per orbit.  The
fraction of initial disk mass that has left the grid through the inner
boundary is given in the final column.  Model SFR has lost the greatest
percentage into the hole, and model SFP the least.  The low mass loss
through the {\it{outer}} grid boundary for all models means that the
radial location of this boundary has had a minimal effect on the overall
simulation.  There are significant differences among the three models in
the amount of gas moving outward, however.  Model SFP has the greatest
relative outflow; by orbit 10, 22\% of the original torus mass has moved
beyond $r=40M$, a radius outside the outer edge of the initial torus.
In model SF0 this amount is 14\%, and in SFR, 10\%.

\begin{table}[ht]
\caption{\label{masstotals}
Mass Budget in Models}
\begin{tabular}{lrrrrl}
 & & & & \\
\hline
Model & $M_0$ & $M_{\rm{final}}$ & $\Delta M_{\rm{in}}$ & $\Delta M_{\rm{out}}$ & $\Delta M_{in}/M_0$  \\
\hline
\hline
SFR        & 0.544   & 0.316   & 0.226  & 0.0065  & 0.41\\
SF0        & 1.000   & 0.675   & 0.308  & 0.0246  & 0.31\\
SFP        & 1.685   & 1.323   & 0.305  & 0.0653  & 0.18\\
SFR-2D     & 0.544   & 0.338   & 0.206  & 0.0060  & 0.38\\
SF0-2D     & 1.000   & 0.709   & 0.295  & 0.0336  & 0.30\\
SFP-2D     & 1.685   & 1.341   & 0.235  & 0.1041  & 0.14\\
SFP-2$\pi$ & 1.685   & 1.302   & 0.324  & 0.0661  & 0.19\\
PN0        & 1.000   & 0.637   & 0.319  & 0.0147  & 0.32\\
\hline
\end{tabular}
\end{table}

The space-time diagram of gas pressure $\langle P_{gas}\rangle$ closely
resembles that of the averaged density (fig.~\ref{rhoav}), and is
therefore not reproduced here.  The pressure decreases with time at the
location of the original pressure maximum as the material in the torus
is redistributed.  The magnetic pressure, $\langle P_{mag}\rangle$,
grows rapidly in the early stages of the simulations ($t < 3$ orbits),
first primarily due to shear of the initial radial field, then due to
the strong growth of the MRI.  Once MRI-driven turbulence is
established, the maximum of magnetic pressure shifts sharply inward
with the infalling magnetized gas.  In models SF0 and SFP the location
of the gas pressure maximum also moves toward the black hole as the
turbulent disk extends inward to the radius of the marginally stable
orbit.

\begin{figure}[ht]
\epsscale{0.26}
    \plotone{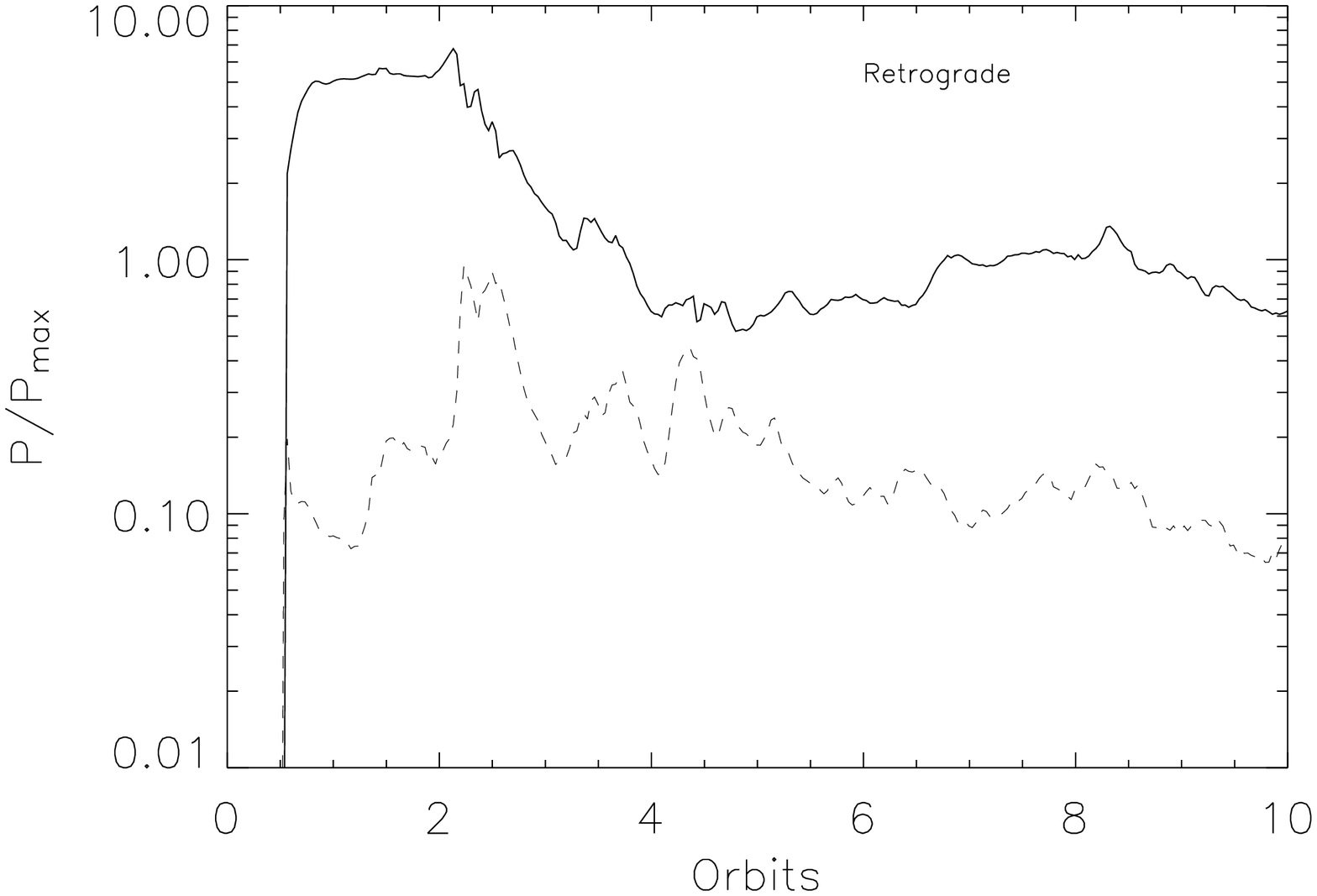}\plotone{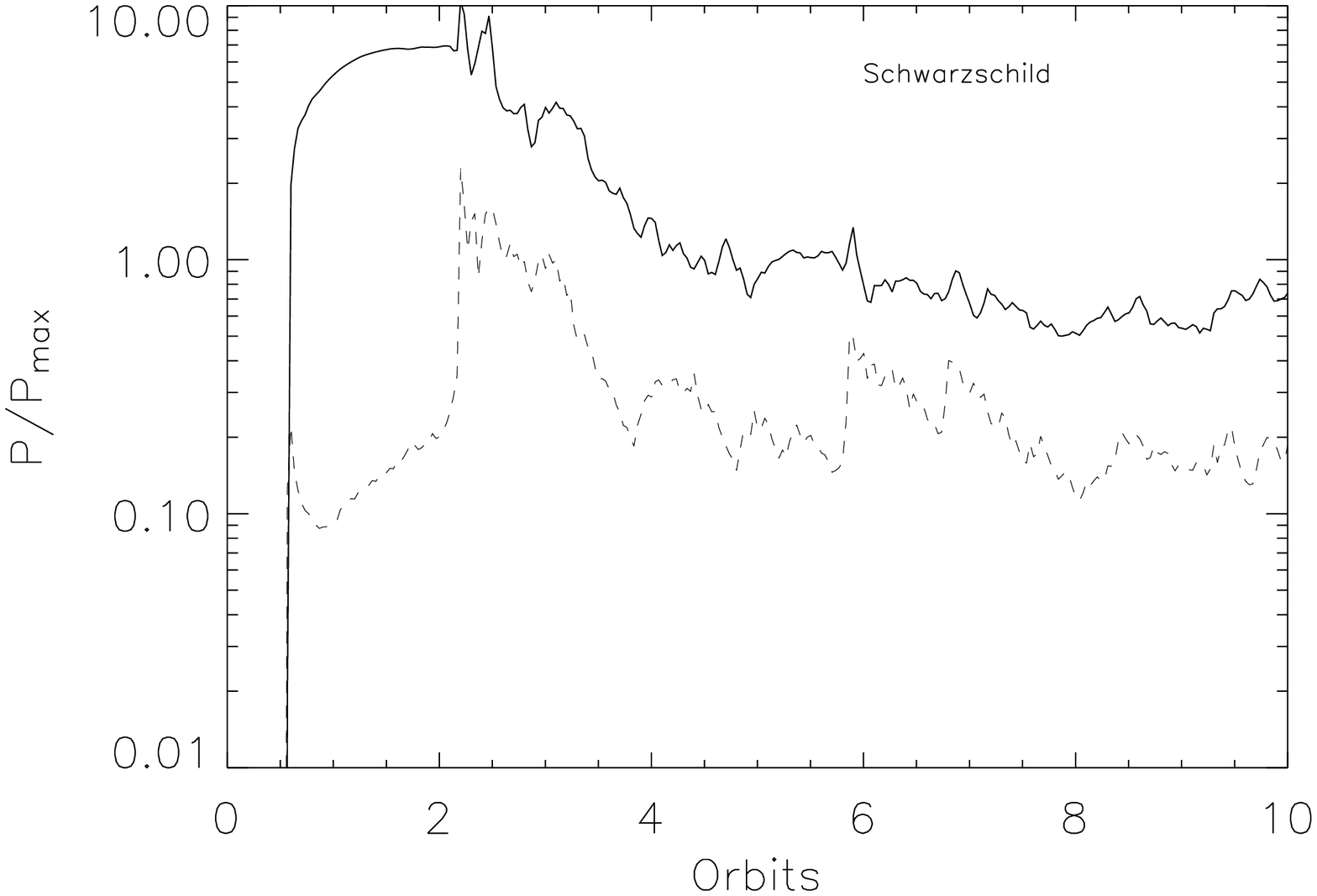}\plotone{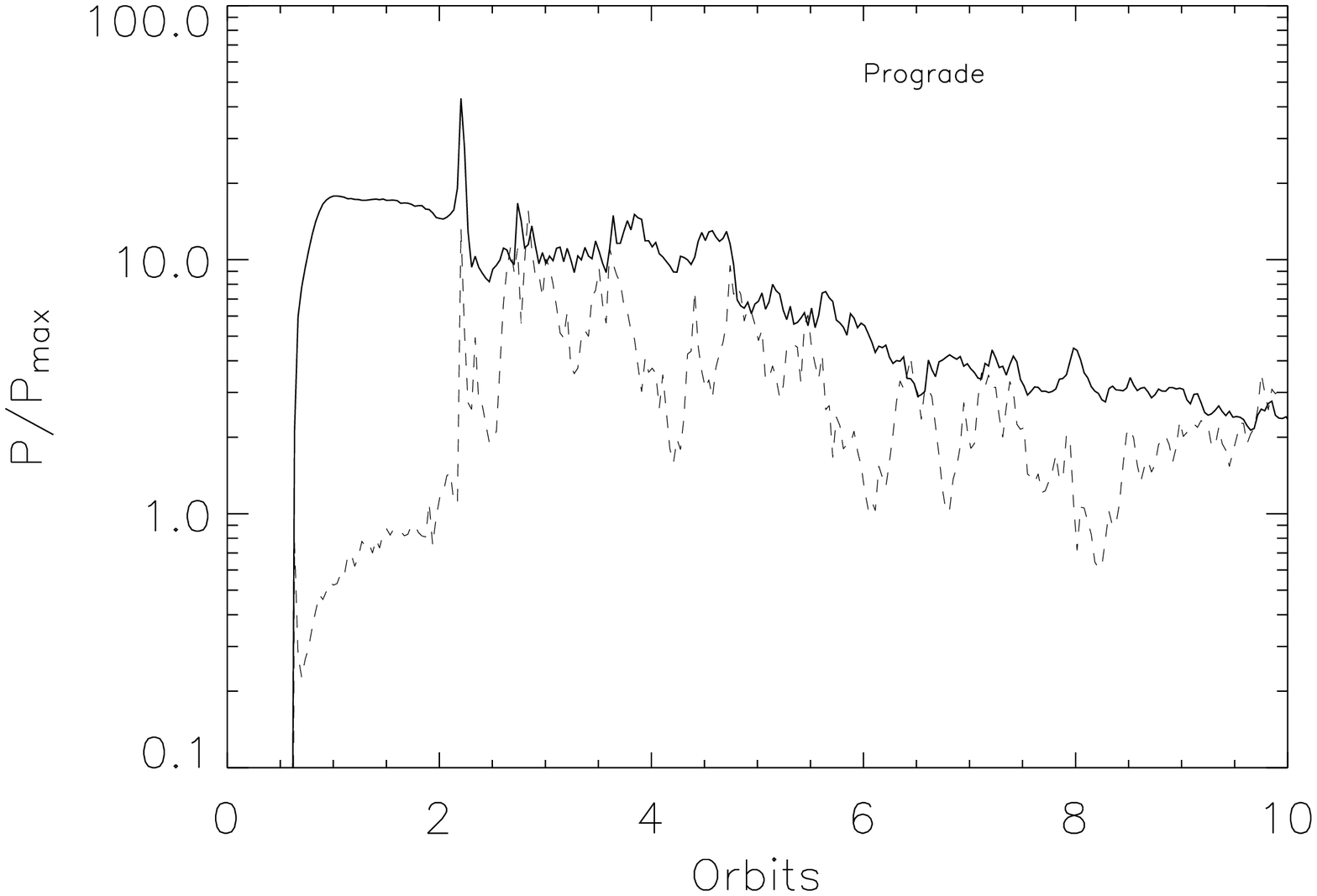}
    \plotone{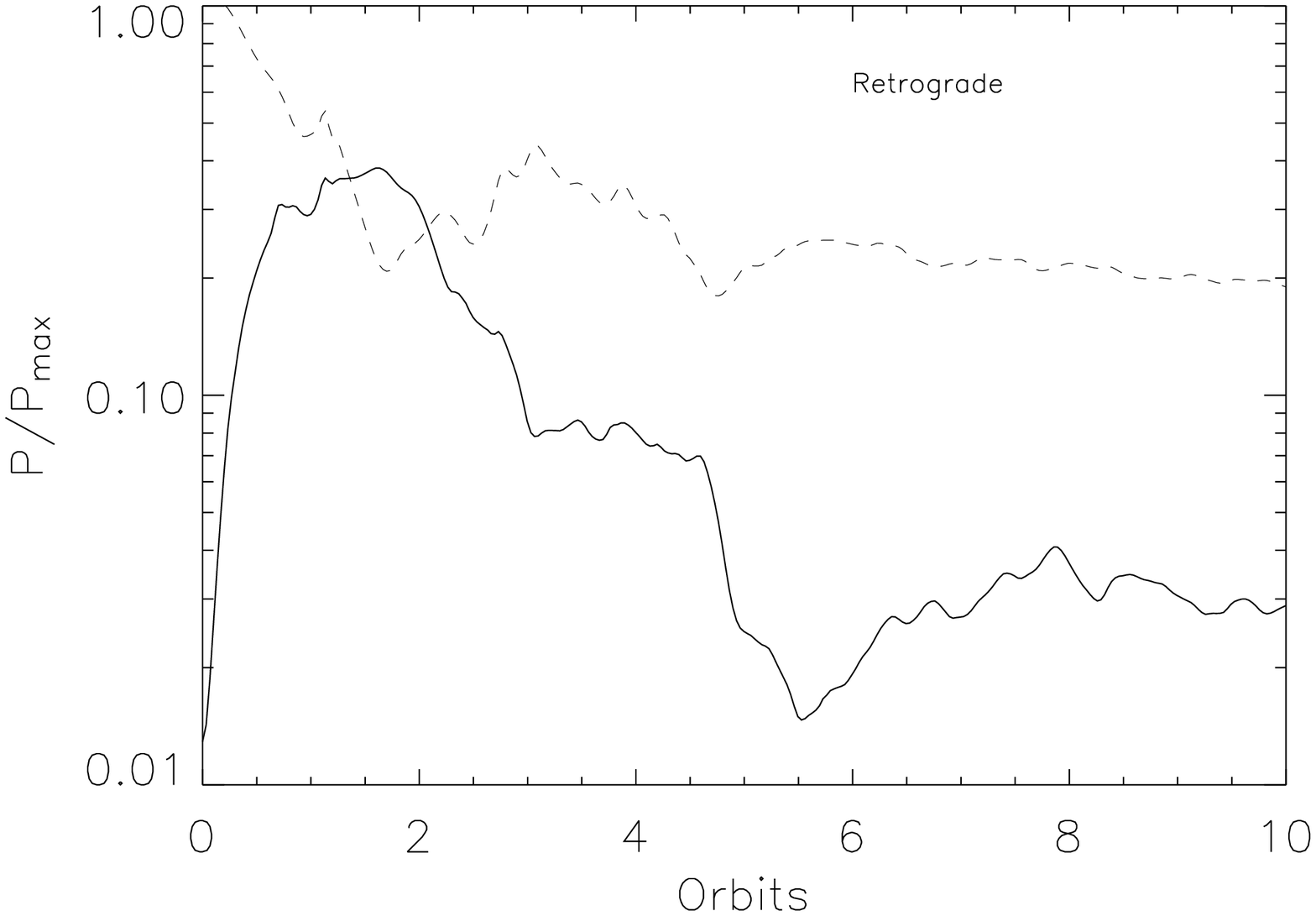}\plotone{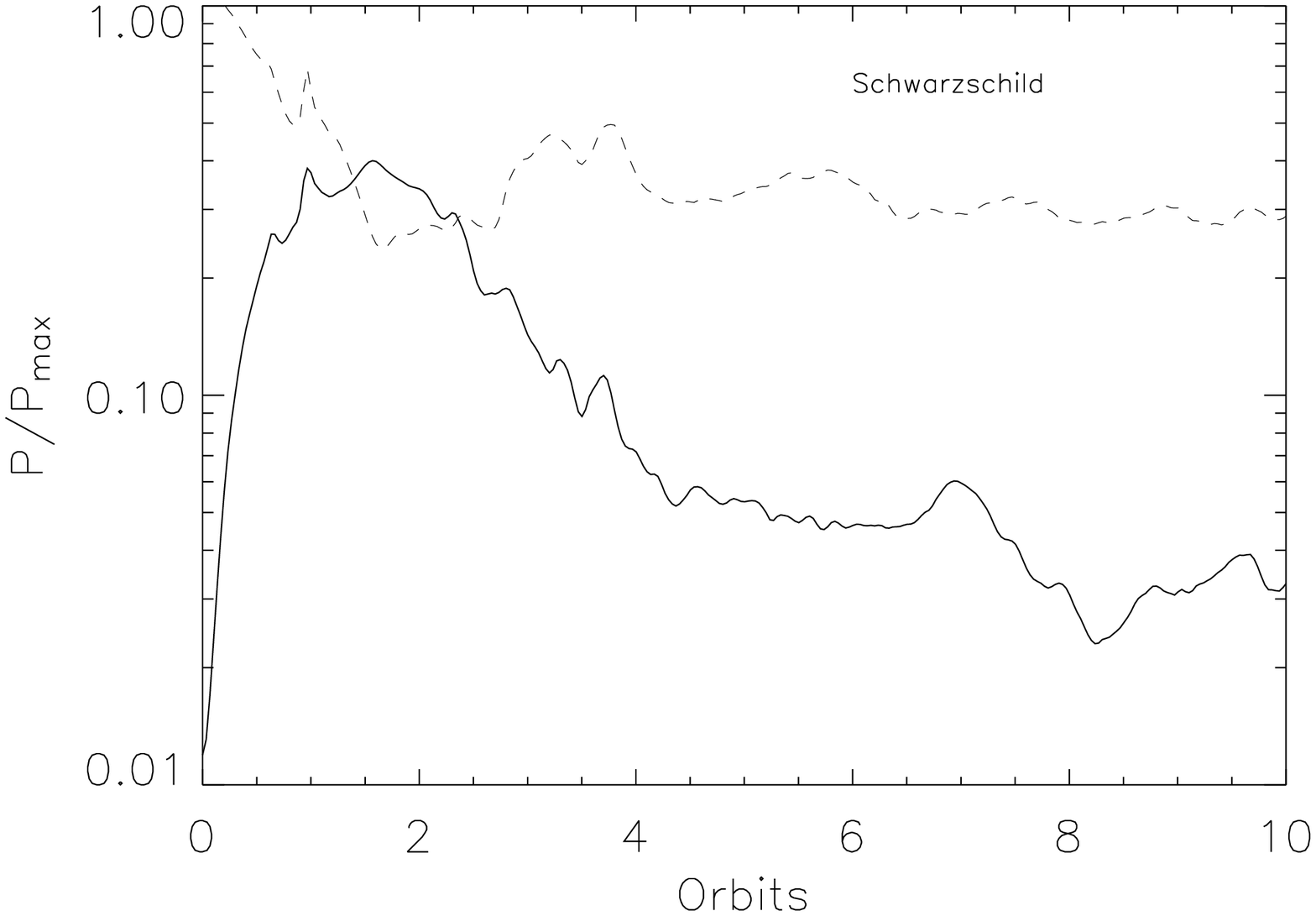}\plotone{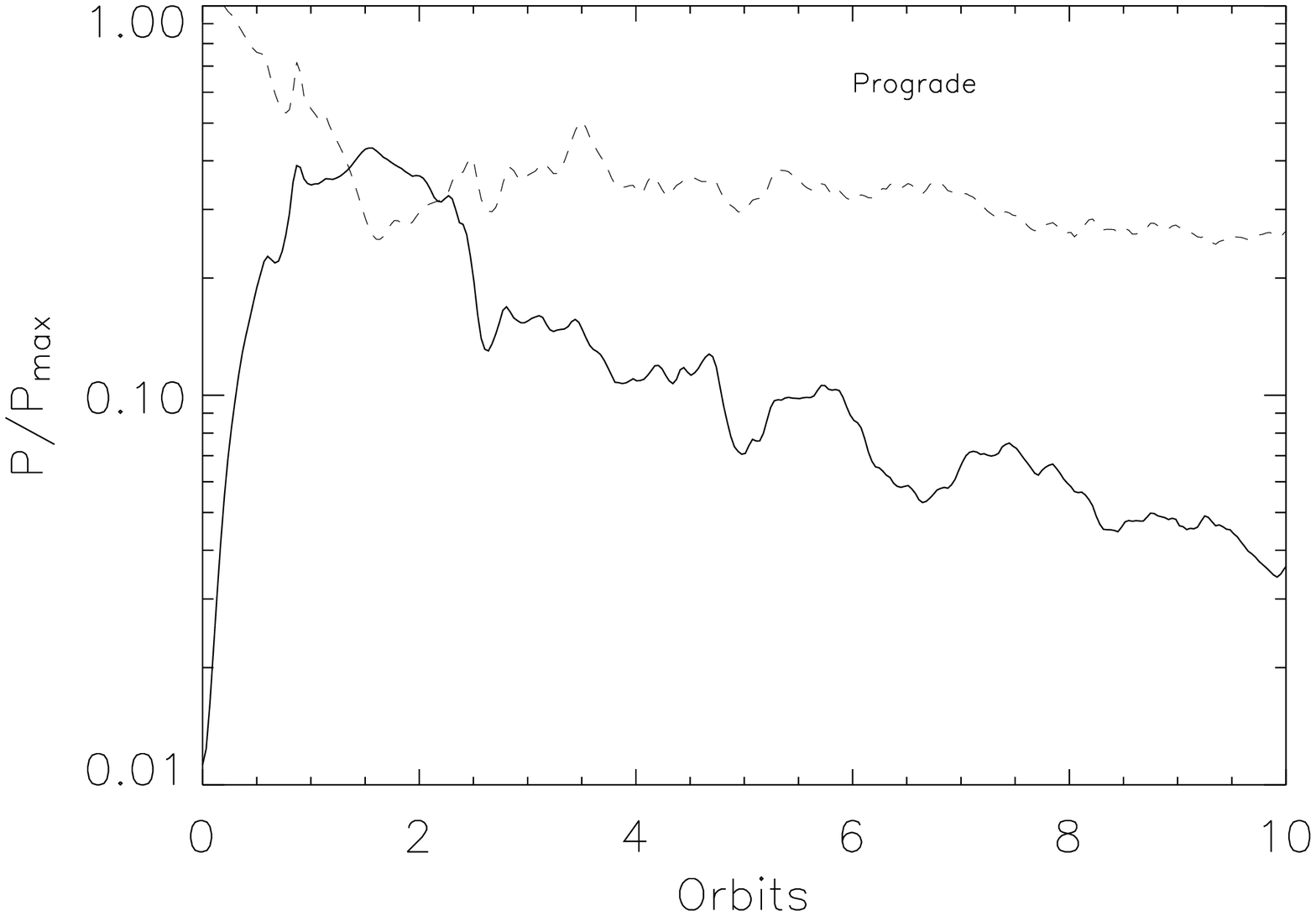}
    \caption{\label{Pressis} 
     Panels (a), (b), and (c) show 
     The time evolution of average
     pressures $\langle P_{mag}\rangle$ (solid
     line) and $\langle P_{gas}\rangle$ (dashed line) near the
     horizon at (a) $r=1.95$ for model SFR, (b) $r=2.33$ for model SF0, and
     (c) $r=1.95$ for model SFP, and inside the torus at the location
     of the initial pressure maximum for (d) SFR, (e) SF0, and (f) SFP.
     The pressures are
     normalized to the initial gas pressure at the pressure maximum 
     and time is in units of orbits at $r_{P\,max}$.} 
\end{figure}

A quantitative view of the pressure evolution in the three models is
given in Figure \ref{Pressis}, where $\langle P_{mag}\rangle$ and
$\langle P_{gas}\rangle$ are plotted as a function of time at the
radius of the initial pressure maximum, and also near the horizon.  In
each case the scale is normalized to the initial pressure maximum.  The
time-evolution inside the torus is very similar in all three runs,
particularly during the initial linear growth phase and nonlinear
saturation.  This indicates that the evolution of the MRI is mostly
independent of the spin of the black hole even relatively close
($r=15M$) to the hole.  Gas pressure dominates except for a brief
period around $t=2.0$ orbits where the violent magnetic transient is at
its peak.  During the extended turbulent phase beyond $t=4$ orbits,
$\beta \approx 5$ to 10.  Near the black hole magnetic pressure dominates,
especially in model SFR.  The two are more nearly equal in model SFP.

\begin{figure}[ht]
    \epsscale{0.35}
    \plotone{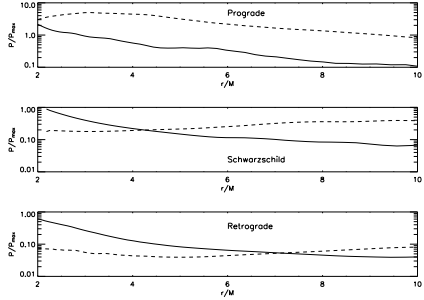}
    \caption{\label{Pinrms} 
     Transition from gas (dashed line) to magnetic pressure (solid line)
     inside the marginally stable orbit is illustrated in models SFP
     (top), SF0 (middle), and SFR (bottom).  The values are
     shell-averages at $t=10.0$ orbits, and normalized to the initial
     torus pressure maximum value.} 
\end{figure}

Where the flow shifts from gas to magnetic pressure dominance is
determined by the location of the marginally stable orbit.  This is
illustrated in Figure \ref{Pinrms}, which shows the averaged gas and
magnetic pressures as a function of radius at orbit 10.  Outside of
$r_{ms}$, the disk evolves due to turbulent stresses driven by the MRI,
and gas pressure continues to dominate.   Inside of $r_{ms}$, the flow
transitions to plunging infall, and the field evolves primarily by flux
freezing.  The magnetic pressure becomes increasingly important,
helping to confine and guide the flow within the plunging region.  The
point where the averaged magnetic pressure exceeds the gas pressure is
located at about $0.7 r_{ms}$ in both SFR and SF0.  In SFP the inner
grid boundary radius exceeds $0.7 r_{ms}$, and the transition to
magnetic dominance is not seen.

\begin{figure}[ht] \epsscale{0.26}
   \plotone{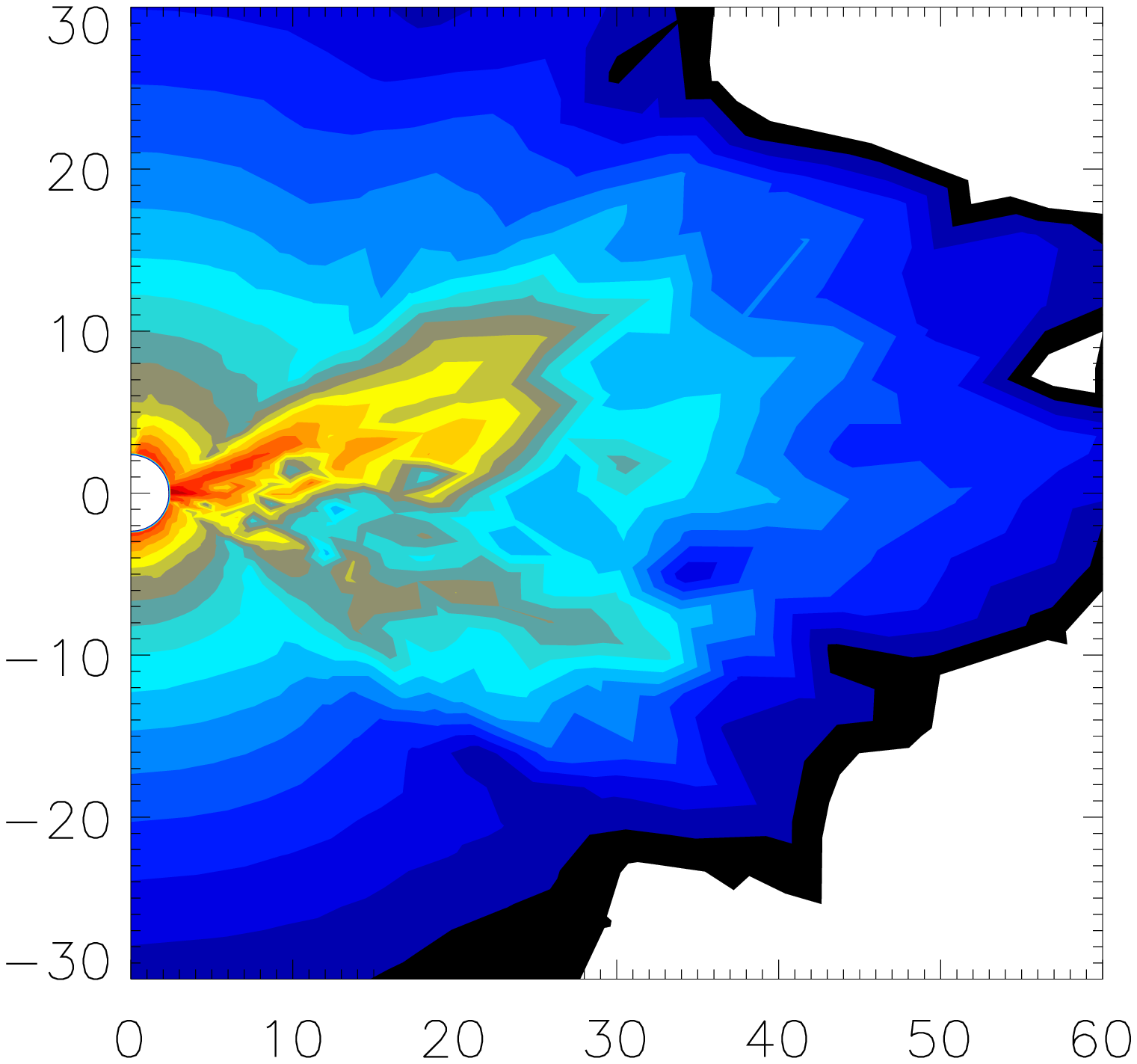}\plotone{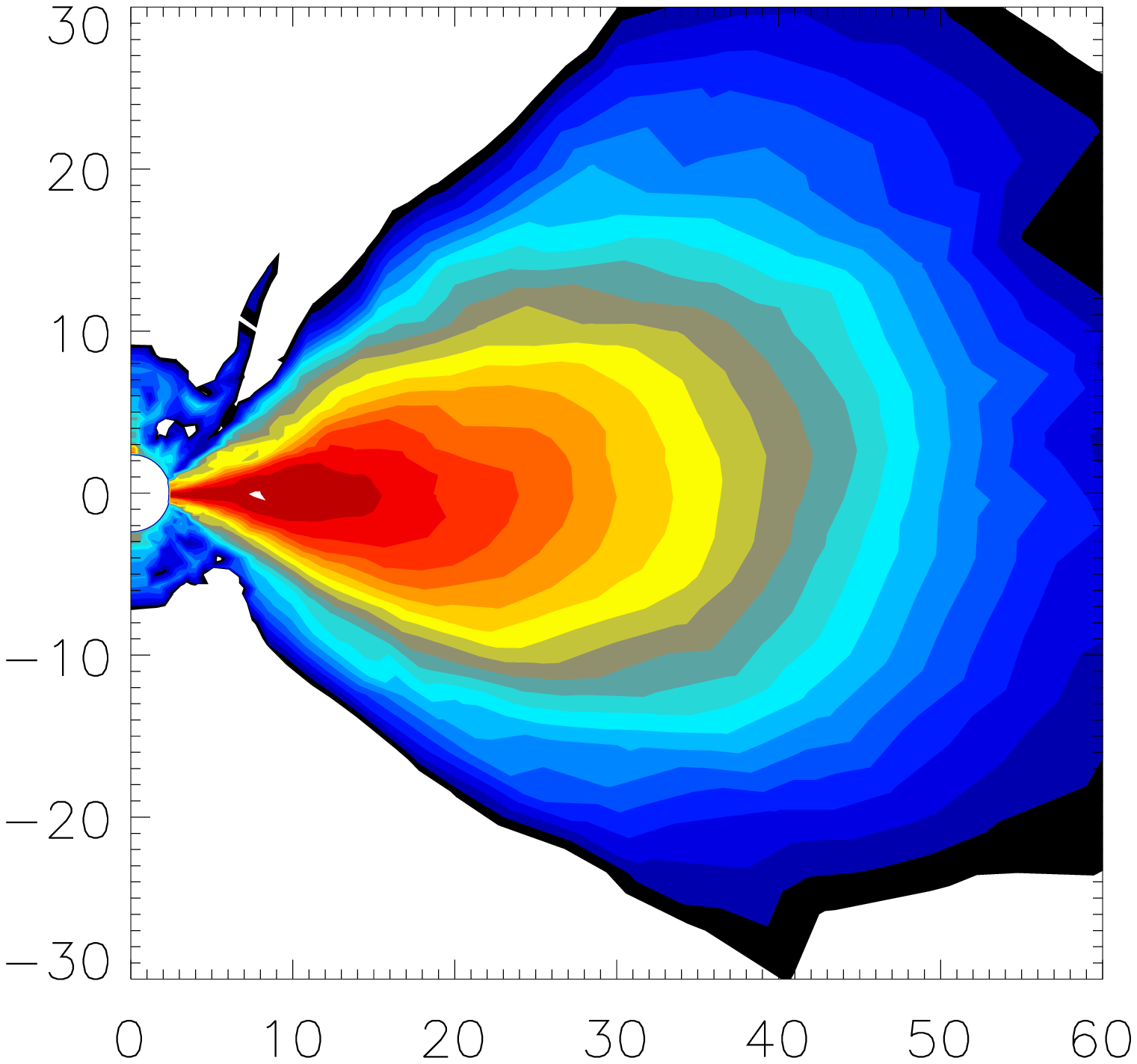}\plotone{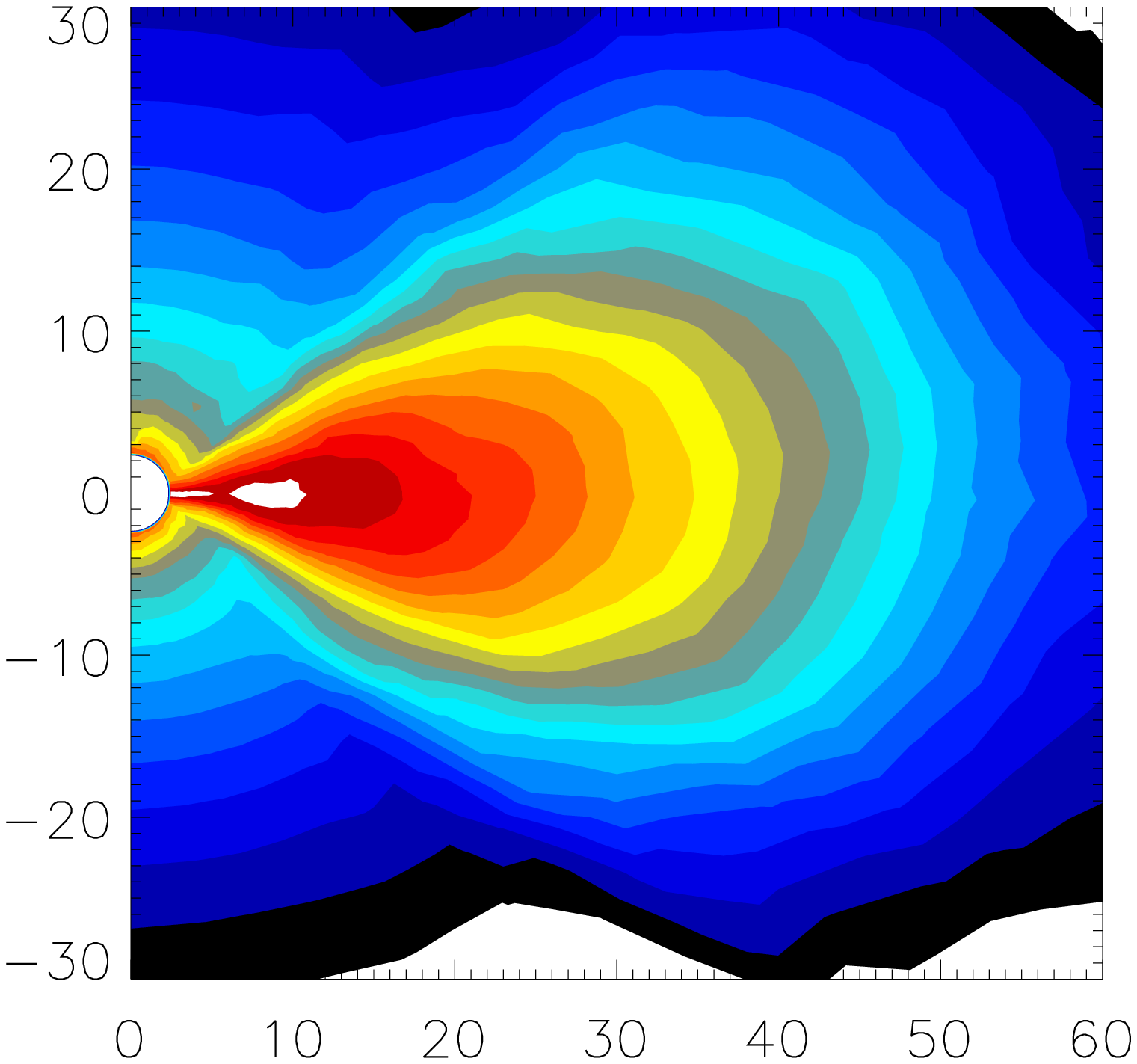}
   \caption{\label{Pazavg} 
     Contour plots of azimuthally-averaged (a) magnetic, (b) gas, and (c)
     total pressure for model SF0 (log scale). 
     }
\end{figure}

Figure \ref{Pazavg} shows the azimuthally averaged gas pressure,
$P_{gas}$, magnetic pressure, $P_{mag}$, and total pressure, $P_{tot}$,
at $t=10.0$ orbits for model SF0 (the other two models have similar
profiles).  Within the disk, total pressure is relatively smooth,
consistent with a thickened equilibrium torus, and dominated by gas
pressure.  Magnetic pressure is increasingly important at the inner
edge of the disk, and gas and magnetic pressure are comparable in the
low-density coronal outflow region surrounding the denser disk.
Magnetic pressure dominates in the funnel region near the axis, the
pressure contours are spherically symmetric, and the magnetic field is
mainly radial.  The material in the funnel consists of fast
outward-moving gas without angular momentum, several orders of
magnitude lower in density than the maximum density in the disk (and
hence below the cut-off in the contour plots shown above).  It is
important to note that the funnel region is initialized to a numerical
vacuum, i.e. the numerical density floor of the code, and never rises
much above this value, so  conclusions about the physical properties of
the funnel region must be reached cautiously.  The point is that
material from the disk, which contains significant angular momentum, is
kept well away from the funnel by the centrifugal barrier.

\begin{figure}[ht]
    \epsscale{0.4}
    \plotone{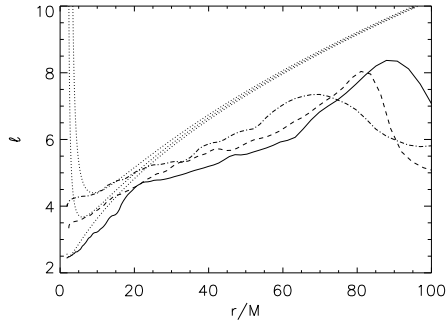}
    \caption{\label{lavg} 
     Specific angular momentum distribution, $\langle l\rangle$
     for models SFR (dot-dashed line), SF0 (dashed line) and SFP (solid
     line ) at $t=10$ orbits.  The Keplerian distributions
     of angular momentum for all three models are shown as dotted lines.} 
\end{figure}

\begin{figure}[ht]
    \epsscale{0.4}
    \plotone{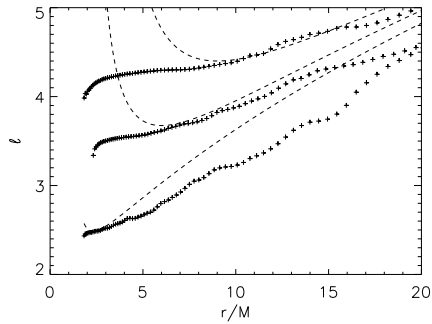}
    \caption{\label{lavgdetail} 
     Specific angular momentum distribution, $\langle l\rangle$, in the 
     inner regions of the flows in the three fiducial models. 
     The Keplerian distribution of angular momentum is shown as a
     dotted line; from top to bottom these curves correspond to the
     retrograde, Schwarzschild, and prograde models.  The crosses
     indicate the values in the models at the radial grid zone locations. } 
\end{figure}

Accretion is driven by the outward transport of angular momentum
through the disk via the MRI.  Each torus began with constant specific
angular momentum, $l$, but  Figure \ref{lavg} shows that by orbit 10 the
distribution $\langle l\rangle (r)$ is approaching a more Keplerian
slope.  For reference, the Keplerian distributions of $l$
are shown as a dotted lines.  In each model the inner portion
of the torus becomes nearly Keplerian after only one orbit of time.
The outer regions evolve more slowly.  The specific angular momentum at
late time is everywhere sub-Keplerian, indicating that some radial
support is supplied by gas pressure (e.g.  fig. \ref{Pazavg}).  The
inflow toward the horizon shows a steady decline in $\langle l\rangle$
down to edge of the grid.  This is indicative of continued magnetic
stress acting on the fluid, a result first noted in the first
pseudo-Newtonian simulations (e.g., Figure 6 of H00).   In the fully
relativistic models SFR and SF0 there is a noticeable roll-off in
specific angular momentum for $r < 3\,M$ (see Fig.~\ref{lavgdetail}).
This roll-off is especially pronounced in the first orbit during the
arrival of low angular momentum fluid that is released from the surface
of the disk during a start-up transient.  The feature remains visible
and well resolved in the late stages as well.  Model SFP does not show
this abrupt drop in $\langle l\rangle$,  and this is likely due to the
lack of an extended plunging region between the marginally stable orbit
and the grid inner boundary.

\begin{figure}[ht]
    \epsscale{0.4}
    \plotone{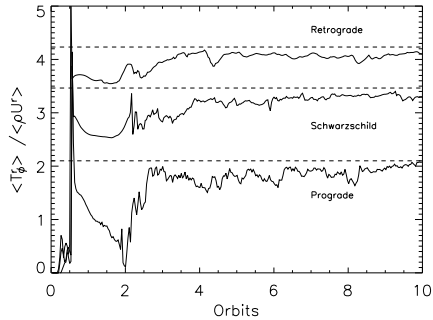}
    \caption{\label{ldot} 
     Total specific angular momentum flux, 
     $\langle{{T^r}_\phi}\rangle / \langle -\rho U^r\rangle$,
     into the black hole for models SFR, SF0, and SFP 
     at $t=10$ orbits.  The value of $(U_\phi)_{ms}$ for a circular orbit
     at $r_{ms}$ is shown for all three models as a dotted line.} 
\end{figure}

A question of particular physical interest is, what is the total
specific angular momentum carried by the accretion flow into the black
hole?  To answer this we divide the total angular momentum flux,
$\langle {T^r}_\phi\rangle$, by the mass flux, $\langle \rho U^r
\rangle$, and examine the value at the inner radial boundary as a
function of time.  This is plotted in Figure \ref{ldot}.  The values
obtained by time-averages over the last three orbits are $-4.06$ for
SFR, $3.32$ for SF0, and $1.94$ for SFP.  These are all slightly below
the value of $U_\phi$ corresponding to a circular orbit at $r_{ms}$
(shown as dashed lines in the figure), and considerably below the value
in the initial torus.

A related question is the value of the specific energy carried into the
hole.  This is computed by the ratio of the energy flux to the mass
flux, $\langle {T^r}_t\rangle/\langle \rho U^r\rangle$ at the inner
radial boundary.  In SFR this value is $-0.969$, in SF0 it is $-0.957$,
and in SFP it is $-0.867$.  These are slightly less bound than the
binding energy of the marginally stable circular orbit, $(U_t)_{ms}$,
although, in all cases, more bound than the value of the initial torus
($-0.98$).  Although we do not reproduce the plot here, the specific
binding energy of the inflow into the hole in SFR and SF0 does not vary
much with time after about 4 orbits, while the value in SFP exhibits
significant fluctuations.  This is another consequence of the absence of
an established plunging inflow inside of $r_{ms}$ for SFP.

\begin{figure}[ht]
    \epsscale{0.5}
    \plotone{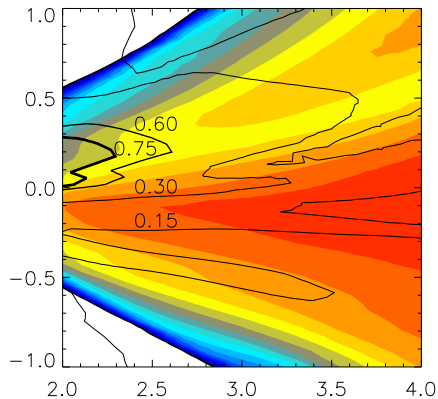}
    \caption{\label{ellaz} 
     Azimuthally-averaged contour plot of specific angular momentum 
     $\langle l\rangle$ in the inner regions 
     of the flow for model SF0. Selected contours of magnetic
     pressure are also shown.  The magnetic pressure contour levels are
     labeled (as fraction of maximum); contour thickness is an indication
     of magnitude of $\|b^2\|$.} 
\end{figure}

The relationship between the magnetic field and the fluid's specific
angular momentum is illustrated in Figure \ref{ellaz}, which is an
azimuthally-averaged plot of the specific angular momentum in the
accretion flow near the black hole for model SF0.  Selected contours of
azimuthally-averaged magnetic pressure are overlaid on the plot.  The
specific angular momentum shows considerable variation with angle
$\theta$.  Regions of high magnetic pressure correspond to regions of
low specific angular momentum.  The fluid with the highest specific
angular momentum that reaches the black hole does so in a region of low
magnetic pressure.  The density-weighted, shell-averaged value $\langle
l \rangle$ cannot completely capture this complexity, but it is clear
from examining points at constant $r$ near the inner edge of the flow,
that it does fairly characterize the reduction in $l$ that occurs near
the horizon.

\begin{figure}[ht]
    \epsscale{0.5}
    \plotone{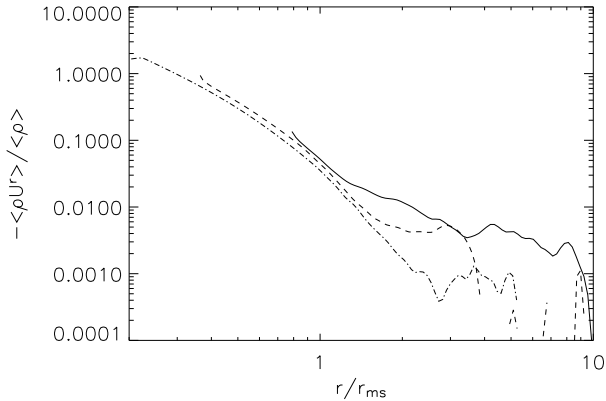}
    \caption{\label{velplt}
     Density-weighted radial velocity average,  $\langle -\rho
     U^r\rangle/\langle \rho \rangle$, as a function of $r/r_{ms}$
     in SFR (dot-dashed line), SF0 (dashed line), and SFP (solid line)
     at $t=10$ orbits.  The graph indicates where, in relation to the
     marginally stable orbit, the flow begins its transition from 
     turbulent disk to streaming inflow.
     } 
\end{figure}

Much of the behavior of the three simulations depends on the location
of the marginally stable orbit relative to the accretion torus.  Krolik
\& Hawley (2002) defined the ``turbulence edge'' of a disk as the point
where the magnetic field switches from being controlled by MHD
turbulence to simple flux-freezing.  Concomitant with this is a
transition from a region where the velocity field is dominated by
fluctuations to one where the velocity corresponds to streaming
inflow.  In the pseudo-Newtonian simulations, Krolik \& Hawley (2002)
found this edge to be located at about $1.2\,r_{ms}$.
Figure~\ref{velplt} is a density-weighted radial inflow velocity
average, $\langle -\rho U^r\rangle/\langle \rho \rangle$, as a function
of $r/r_{ms}$ at $t=10$ orbits in the three fiducial models.  The
shell-averaged velocities within the turbulent portions of the disk are
relatively small and variable.   At a point just outside of $r_{ms}$,
however, $U^r$ begins a smooth and steady increase with decreasing
radius.  From $r_{ms}$ inward the slopes of the three curves show close
agreement.

\subsection{Comparisons:  Axisymmetric and full $2\pi$ Evolution}

\subsubsection{Axisymmetry}

Reducing the extent of the azimuthal direction is a very effective way
of reducing the computational demands of a given problem.  Axisymmetric
(two-dimensional) simulations are particularly useful in this regard.  But
to what extent do axisymmetric simulations convey useful information?  The
limitations of axisymmetric pseudo-Newtonian simulations have been
previously discussed by Hawley, et al.~(2001).  Here we carry
out axisymmetric versions of our three fiducial models to investigate
these issues in the general relativistic regime.  These are designated
SFR-2D, SF0-2D, and SFP-2D.  The axisymmetric
simulations will also provide a more complete picture of the role of
azimuthal modes in the growth and maintenance of the MRI in the Kerr
metric.

\begin{figure}[ht]
     \epsscale{0.4}
     \plotone{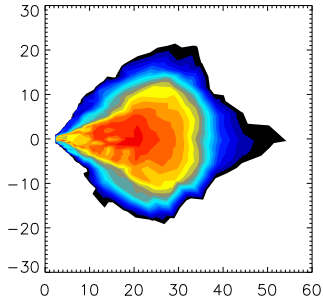}
     \caption{\label{2DMRI} 
     Plot of log density $\rho$ for model SF0-2D at $t=3.0$ orbits.
     The axisymmetric MRI operating on vertical field (the ``channel 
     solution'') leads to the long linear features seen here.}
\end{figure}

The axisymmetric simulations are run for a time equivalent to $10$
orbits at the pressure maximum.  As in the fiducial simulations, the
magnetic pressure increases rapidly due to the growth of toroidal field
by shear.  The toroidal field MRI, however, is inherently
nonaxisymmetric, so only the poloidal MRI modes operate.  As has
previously been noted (e.g., Hawley et al.~2001), one of the main
consequences of axisymmetry is that the evolution is dominated by a
particularly violent form of the poloidal MRI, the so-called ``channel
solution'' (Hawley \& Balbus 1992).  Figure \ref{2DMRI} shows the
density for model SF0-2D at $t=3.0$ orbits (models SFR-2D and SFP-2D are
similar).  The prominent radially-extended features are characteristic
of the channel solution.  The channel solution is itself subject to an
instability which destroys its coherence, but that instability is
nonaxisymmetric (Goodman \& Xu 1994).  Another known consequence of
axisymmetry is that magnetic turbulence cannot be sustained
indefinitely.  In axisymmetry the toroidal field cannot be converted
into poloidal field to create a dynamo;  this is Cowling's anti-dynamo
theorem.

The combination of the channel solution plus the anti-dynamo theorem
paradoxically produces both more violent initial accretion and weaker
long-term accretion.  In the axisymmetric models the overall accretion
totals are reduced from their three-dimensional counterparts (Table
\ref{masstotals}).  However, in axisymmetry the initial infall of matter
is more pronounced; there are significant bursts of accretion from the
disk to the horizon.  

\begin{figure}[ht] 
\epsscale{0.4} 
\plotone{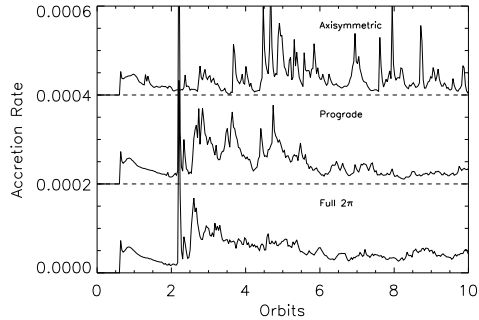}
\caption{\label{mdotcompare}
    Accretion rate, $\dot M$, at the inner radial
    boundary ($r_{min}$) in models SFP-2D, SFP, and SFP-2$\pi$.  
    The accretion rate is normalized by the initial
    torus mass.   The value for SFP-2D (axisymmetric) is offset by 0.0004,
    and SFP (standard Prograde) is offset by 0.0002.
     } 
\end{figure}

\begin{figure}[ht] 
\epsscale{0.4} 
\plotone{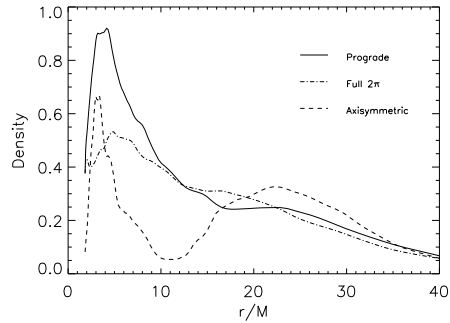}
\caption{\label{rhocompare}
     Averaged density $\langle \rho\rangle$ as a function of radius at
     orbit 10 for models SFP-2D (dashed line), SFP (solid line)
     and SFP-2$\pi$ (dot-dashed line).   The density is
     normalized to the maximum value in the initial torus.
     } 
\end{figure}

Figure \ref{mdotcompare} shows the accretion rate into the hole for
model SFP-2D compared with that in SFP.  The most obvious feature is
that accretion in SFP-2D is highly time-variable, even episodic.  In
addition, the initial accretion flow takes longer to reach the hole;
strong inflow doesn't occur until 4.5 orbits, when there are sharp,
high-amplitude spikes in $\dot M$.  After this, accretion continues at a
steady low rate with periodic brief increases.  In this model, matter
accumulates in a mini-torus near the black hole, much like it does in
three dimensions.  At late time the mini-torus continues to accrete into
the hole, but inflow from the main disk to the mini-torus is greatly
reduced.  Figure \ref{rhocompare} compares the averaged density $\langle
\rho\rangle$ at orbit 10 in model SFP and SFP-2D.  One can see that the
mini-torus is more or less distinct from the main torus, indicating that
the accretion flow between these two regions has been greatly reduced.

In models SFR-2D and SF0-2D (not shown) there are strong
bursts of accretion between orbits 3 and 5, and beyond that time the
accretion rate declines steadily, punctuated by an occasional flare.  By
the end of the three axisymmetric models, the accretion rate has quieted
considerably and much of the matter remains out in the region of the
initial torus, relatively undisturbed.

\subsubsection{Full $2\pi$ plane}

It has been noted in pseudo-Newtonian studies that quarter-$\phi$-plane
simulations are often adequate for capturing the essential aspects of
accretion disk evolution (Hawley 2001; Nelson \& Papaloizou  2003).  
Hawley (2001) found good qualitative agreement between the two types
of grids, with an overall 10\% reduction in field and stress levels
in the restricted-$\phi$ simulations.

To examine the differences between a full- and a quarter-plane
simulation in the present context, we repeat SFP on a $128^3$ grid with
the full range in the $\phi$ coordinate ($0 \le \phi \le 2\,\pi$).
This simulation is labeled SFP-2$\pi$.  Since we simply increase the
number of $\phi$ zones by a factor of 4, the grid zone size, $\Delta
\phi$, remains the same.   Of the models run here, the prograde torus
provided the greatest radial extent of turbulent disk evolution, and
provides the most interesting comparison.  Polar and equatorial slices
through the dataset for the density $\rho$ are similar to Figure
\ref{SFPtwo}, so the gross overall dynamics of the disk are largely
unaffected by the choice of $\phi$ grid.

There are differences, however.  A comparison of the density at orbit 10
in SFP-2$\pi$ and SFP (fig.  \ref{rhocompare}) shows that the density in
the mini-torus is substantially reduced in SFP-2$\pi$.  Although the
mini-torus is initially comparable in size in the two models, it seems
to partially dissipate in the late stages of SFP-2$\pi$.  In SFP the
mini-torus appears to dissipate at $t \approx 5$ orbits, but then
reconstitutes itself.  This same behavior was seen in the
pseudo-Newtonian simulation of Hawley \& Balbus (2002); an inner torus
formed, dissipated, and reformed during the length of the simulation.
In SFP-2$\pi$ the density within the mini-torus declines after its
formation until the end of the run.  How can we account for this?  First,
the mini-torus forms, in part, because the accretion rate into the
near-hole region doesn't necessarily match the accretion flow that
passes beyond the marginally stable orbit and into the hole itself.  The
initial nonlinear MRI, for example, can rapidly dump considerable
material into the near-hole region.  At later times inflow from the
extended disk may be reduced, while an increase in turbulence within
the mini-torus can increase the rate of its accretion into the hole,
thereby yielding the observed descrease in density within the SFP-2$\pi$
mini-torus.  It would seem, therefore, that greater stress levels in
SFP-2$\pi$ have a noticeable impact on disk structure.

The long-term behavior of the accretion rate may constitute another
significant difference between the models.  Figure \ref{mdotcompare}
compares the accretion rates in SFP-2$\pi$ and SFP.  The fluxes are
similar in magnitude, but the full-grid simulation is smoother, and has
a slightly greater mass flux after $t=7$ orbits.  Table \ref{masstotals}
shows that the total integrated accretion in SFP-2$\pi$ is about 6\%
greater than that of SFP.  This would indicate that the low-order
azimuthal modes, whose existence is precluded in the quarter-grid
simulations, do play a role in the accretion process and can increase
the effective stress and angular momentum transport.  The reduction in
the level of the fluctuations, particularly when one also considers the
axisymmetric result, also shows the role of the nonaxisymmetric modes in
decreasing the coherence of the axisymmetric MRI modes (i.e., the
channel solution).  It is also possible that a reduction in $\Delta
\phi$ would similarly reduce the fluctuations in $\dot M$, even with a
quarter-grid domain.  This hypothesis remains to be tested with higher
resolution simulations.

\subsection{Comparison with a Pseudo-Newtonian Simulation}

To compare the results from the relativistic code with those that use a
pseudo-Newtonian potential, we ran model PN0, an analogue of SF0.
These two models have several differences in addition to Newtonian
versus relativistic physics.  The PN0 code is the same as used in
earlier studies (e.g., Hawley \& Krolik 2001, 2002) and, although the
numerics in the two codes are based on the same principles of
time-explicit Eulerian differencing, they are, of course, distinct
algorithms.  The PN0 code employs a cylindrical grid, $(R,\phi,z)$,
with a unit of radius twice that of the Schwarzschild metric, i.e.,
$R=2M$.  The radial grid has 128 zones and is equivalent to that of
SF0, while the $z$ grid has 128 zones, with 64 of the zones located
between $z= \pm 5$.  In PN0 the initial torus has a pressure maximum at
$R=7.65$ and an inner edge at $R=4.75$.  The angular momentum is
$l=3.18$ (unit value smaller by $2^{1/2}$ compared to SF0).  The
initial magnetic field consists of toroidal loops with $\beta = 100$,
set up in the torus according to eqn.~(\ref{vecpot}).  The orbital
period at the pressure maximum is $T_{orb} = 115.6$.  The
pseudo-Newtonian orbital period differs from that of SF0 by the radial
unit difference $2^{3/2}$ ($P_{orb} \propto \Omega^{-1} \propto
r^{3/2}$), plus a factor due to time-dilation, $\propto (1-2/r)$.  In
the GR code time is measured in terms of the clock of the observer at
infinity, whereas there is no distinction between local and global time
in the Newtonian computation.  When comparing the results from the two
simulations we normalize time to orbits at the initial pressure
maximum.

\begin{figure}[ht] 
\epsscale{0.5} 
\plotone{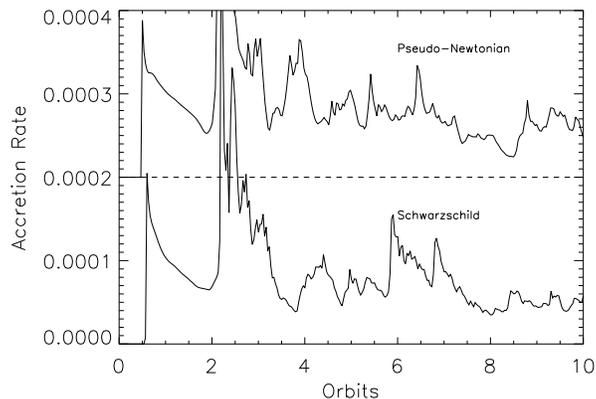}
\caption{\label{pnmdotcompare}
    The accretion rate, $\dot M$, at the 
    inner radial boundary ($r_{min}$) in models SF0  and PN0.
    In both cases the accretion rate is normalized by the initial
    torus mass,  and time is in units of 
     the orbital period at the initial torus pressure maximum.
     The value for PN0 (pseudo-Newtonian) is offset by 0.0002.
     } 
\end{figure}

Figure \ref{pnmdotcompare} compares the history of the accretion into
the hole for runs PN0 and SF0.  The plots are very similar in overall
shape.  In both models the initial phase of magnetic pressure-driven
accretion, the onset of the MRI saturation phase, and the transition to
more or less steady turbulent accretion occur at the same time,
normalized by the orbital period at the pressure maximum.
Table \ref{masstotals} gives the mass budget for model PN0; the values
are normalized to the initial torus mass.  The mass loss through the
outer boundary includes both the upper and lower $z$ boundaries as well
as the outer cylindrical radial boundary.  
PN0 has lost a comparable amount of mass as the
relativistic SF0 model.

\begin{figure}[ht]
    \epsscale{0.5}
    \plotone{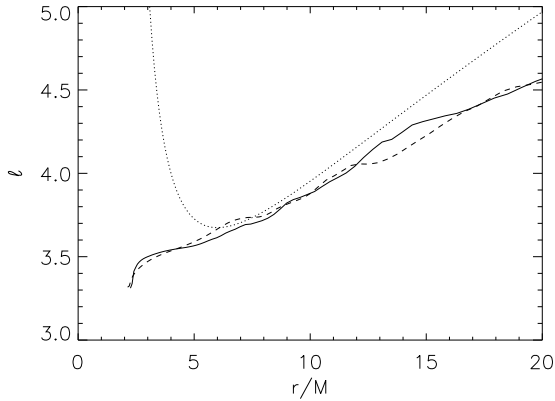}
    \caption{\label{pnelcompare} 
     Specific angular momentum distribution, $\langle l\rangle$,
     for models SF0 (solid line), and PN0 (dashed line)
     at $t=10$ orbits.  The Keplerian distribution
     of angular momentum for SF0 is shown as a dotted line.} 
\end{figure}

Figure \ref{pnelcompare} shows the averaged specific angular momentum 
$\langle l \rangle$ at 10 orbits in models SF0 and PN0.  Note that in 
PN0 the average is taken over cylinders rather than spherical
shells.  However, since it is a density-weighted average, and most
of the mass is near the equator, this difference should not be too
great.  One sees that there is substantial agreement between the two
curves, including a continued decline inside of the marginally
stable orbit.  PN0 also shows a slight downturn in slope near the
horizon, although not as pronounced as in SF0.

One can compute a mass averaged inflow velocity by dividing $\dot M$ by
$\langle \rho \rangle$.  This averaged inflow velocity at 10 orbits in
PN0 is larger than in SF0 by about 20\% at $r_{ms}$, and increases more
rapidly inward.  The PN0 velocity reaches a maximum of about 2.4$c$ at
the horizon while SF0 remains subluminal and approaches $c$.
If accurate velocities are important to the analysis,
a relativistic treatment is clearly preferred.

\section{Discussion}

In this paper, we present a preliminary survey of the properties of
accreting tori in the Kerr metric.  This work is the relativistic
analogue of earlier work by Hawley (2000) which used a pseudo-Newtonian
potential.  We carried out three-dimensional simulations of an
accretion torus around prograde and retrograde Kerr holes with large
$a$ values, as well as around a nonrotating Schwarzschild hole.  These
simulations are contrasted with axisymmetric simulations and a
pseudo-Newtonian simulation with equivalent initial conditions.

In comparing the three fiducial models, we find that the most important
effect of black hole rotation is in setting the location of the
marginally stable orbit.  In retrograde model SFR, which has the
outermost $r_{ms}$, the initial torus begins with its inner edge almost
at the marginally stable orbit.  As pressure builds and angular momentum
is transported out through the torus, a slender accretion flow is set up
which remains relatively smooth in the late stages of the simulation.
The ease with which the matter is accreted is indicated by the
observation that over the course of the simulation, $42\,\%$ of the
initial disk mass enters the black hole in 10 orbits.  At the other
extreme, in the prograde model, SFP, the gas must slowly accrete
from the initial torus down to $r_{ms}=2.32\,M$, a radius just outside
the static limit.  This accretion is driven by angular momentum
transport from the MHD turbulence, and the flow develops into a dense,
slightly sub-Keplerian disk.  The inner edge of this disk features a
local pressure maximum inside $r \approx 5\,M$ which we refer to as the
``mini-torus.'' All of these factors reduce the total accretion through
the inner radial boundary into the hole over the course of the 10 orbit
evolution.  Only 18\% of the initial torus mass is lost into the hole.
A greater fraction of the mass is lifted to radii outside the outer
boundary of the initial torus.  The Schwarzschild model SF0 is an
intermediate case which develops both a thick, turbulent accretion flow
and a plunging inflow.  SF0 loses $31\,\%$ of its initial mass to the
black hole.

In all three models the accretion rate into the hole is highly time
variable.  SFR shows less variability on short time scales.  The
accretion rate into the hole is determined by the rate at which
material from the turbulent disk is fed into the plunging flow from the
disk's ``turbulence edge,'' a point located near, but generally
slightly outside, the radius of the marginally stable orbit.  The
accretion rate variability will be roughly determined, therefore, by
the orbital frequency at the turbulence edge.  Assuming that
fluctuations in $\dot M$ have observable consequences (e.g., as a
possible origin of high-frequency QPOs), then these simulations
confirm the idea that the hole's spin parameter $a$ helps determine
the variability frequency.  The intrinsic variability of the
turbulence itself, however, prevents the relationship between $a$ and
variability frequency from ever being more than approximate.

As in our earlier hydrodynamic simulations, we observe the effect of
frame dragging most prominently in the retrograde model SFR, where the
spiral accretion stream shows the characteristic ``dog leg'' pattern
near the hole as the stream is dragged into co-rotation with the black
hole inside $r \approx 3\,M$.  The effects of prograde frame dragging
are much more difficult to discern directly.

We find that, as in earlier pseudo-Newtonian simulations, an extended
magnetized corona forms where the magnetic pressure is comparable to
the gas pressure.  Much of this corona is flowing radially outward.  
The largest outflows occur in SFP; again, because it has higher
specific angular momentum relative to the marginally stable orbit,
the gas is not as easily lost to the hole as in SFR.

In all the models the initial magnetic field corresponded to a
volume-averaged strength of $\beta = 100$.  In the subsequent evolution
the disk remains dominated by gas pressure, with $\beta \sim 10$, while the
coronal regions are more near equipartition, with $\beta \sim 1$.  Magnetic
pressure dominates within the centrifugal funnel region near the black
hole.  However, there is no significant matter density within the
funnel; the specific angular momentum of the gas prevents the accreted
gas from approaching the axis.

An observation that clearly emerges from the simulations is the
dominant role of the magnetic fields within the plunging inflow inside
the marginally stable orbit.  In models SF0 and SFR there is a
noticeable transition from a gas-pressure dominated to a magnetic-pressure 
dominated flow inside $r \approx 0.7\,r_{ms}$.  For model SFP,
however, this transition does not occur; the marginally stable orbit
lies too close to the inner radial boundary.  Instead we see
near-equality of the two pressures at the inner radial boundary.  A
higher-resolution simulation of model SFP which includes  more zones
within the ergosphere is required to resolve the plunging inflow.

The magnetic fields within the plunging region continue to exert a
stress on the fluid.  Regions of strong field correspond to the
lowest specific $l$ values in the gas entering the hole.  In all
three models the specific angular momentum flux entering the hole, 
$\langle {T^r}_\phi \rangle/\dot M$, is slightly less than the value
associated with the marginally stable orbit:  96\% for SF0 and SFR,
and 92 \% for SFP.


We also compared the fiducial simulations to axisymmetric simulations.
The overall initial evolution of the axisymmetric and three-dimensional
simulations are similar.  The magnetic pressure increases, causing the
torus to expand and creating outflows, while the angular momentum
within the torus is redistributed to a more Keplerian form.  However,
the duration of MRI-driven turbulence is short-lived in the
axisymmetric case since the azimuthal modes that sustain the MRI are
absent under axisymmetry.  Further, as the SFP-2D simulation
demonstrates, the turbulence can die out in one region of the flow
while continuing in others, therefore significantly altering the global
evolutionary properties.

As noted earlier by Hawley \& Balbus (2002), axisymmetric simulations
tend to be dominated by the poloidal field MRI channel solution which
is characterized by radially extended filamentary structure and large
angular momentum transport.  The channel solution is also present in
the three-dimensional simulations, but it is short-lived due to
nonaxisymmetric modes, and its breakup heralds the onset of
MRI-driven turbulence.  These axisymmetric MRI modes are highly 
time-variable, causing accretion to occur in bursts.  Comparing the results
of the full $2\pi$ plane simulation SFP-2$\pi$ with the standard
quarter-plane simulation shows that with greater axial extent, the
accretion flow into the hole becomes smoother in time.  This points to
the role of nonaxisymmetric modes in moderating the influence of the
axisymmetric channel solution.

Since axisymmetric simulations cannot model essential aspects of the
accretion process, their utility is somewhat limited.  Used with care
they can provide an inexpensive way to evolve an idealized initial
condition, such as a constant-$l$ torus with loops of field, into a more
extended near-equilibrium structure that then serves as the starting
point for a full three-dimensional simulation.  They also provide a
valuable tool for surveys of various initial conditions.  Since
axisymmetric simulations can be much more highly resolved, they also
should continue to be useful as numerical experiments to investigate
specific physical processes in detail.

In comparing full- and quarter-plane simulations we find that, while
there are noticeable differences, the quarter-plane simulations capture
much of the qualitative behavior.  This is the consistent with the
conclusion from previous pseudo-Newtonian studies.  Their affordability
makes them very useful for running a range of models, such as the
present work.  The accretion rate in the full-plane simulation is less
strongly variable, presumably due to a reduction in the power of the
axisymmetric poloidal field MRI.  The time-averaged accretion rate,
however, is higher, and this influences the size of the mini-torus that
forms in the prograde runs.  The optimal strategy may be to run models
on restricted grids until the startup transients are past and the
turbulent disk is well-established.  The $\phi$-grid can then be
increased to get an improved representation of the approximately
steady-state disk.

A direct comparison of relativistic model SF0 and pseudo-Newtonian
model PN0 reveals good agreement:  the two simulations show similar
late-time angular momentum distributions as well as comparable accretion
histories through the inner radial boundary.  In the regions outside the
marginally stable orbit relativistic and pseudo-Newtonian results
overlap, which should not be surprising.  This is a region of weaker
gravitational field and the pseudo-Newtonian potential is a very good
approximation to the Schwarzschild metric.  The differences between the
two simulations arise in the region inside $r \approx 3M$, where general
relativistic effects become prominent.  This is also the region where
the velocities in pseudo-Newtonian models are likely to be the most
inaccurate.  In relativity velocities are limited to less than $c$,
whereas there is no such limit in a pseudo-Newtonian simulation.  We
further note that there is no unique time-normalization that is
appropriate for all locations in the flow when comparing a relativistic
simulation with a Newtonian one.  If details of the time-variability of
the flow in the near hole region are important, relativistic
calculations are required.  Simulating accretion into Kerr holes also
requires a relativistic treatment.

As the time-dependent models improve, it is important to connect the
results to observable phenomena.  Recently, Armitage \& Reynolds (2003)
took a first step toward this goal with a study of the effects of time
variability on certain emission processes in the inner regions of a
simulated black hole accretion disk.  Machida \& Matsumoto (2003) have
evolved a global disk to search for regions within the plunging flow
that might account for X-ray flares.  These studies must ultimately be
done with full relativity.  In the present work we see both qualitative
and quantitative differences among the models with different $a$ values
in the plunging region inside the marginally stable orbit, where effects
of black hole rotation are increasingly important.  More detailed
analyses of this plunging region will be the subject of subsequent
papers.

\acknowledgements{This work was supported by NSF grant AST-0070979 and
NASA grant NAG5-9266.  The simulations were carried out on the Origin
2000 system at NCSA, and the Bluehorizon system of NPACI.}



\begin{thebibliography}

\bibitem[Armitage, Reynolds, \& Chiang (2001)]{arc01} Armitage, P. J., 
Reynolds, C. S., \& Chiang, J.  2001, ApJ, 548, 868
\bibitem[Armitage \& Reynolds (2003)]{ar03} Armitage, P., \& Reynolds, C. 2003,
MNRAS, in press
\bibitem[Balbus \& Hawley (1991)]{BH:91} Balbus, S.~A., \& Hawley, J.
~F. 1991, ApJ, 376, 214
\bibitem[Balbus \& Hawley (1992)]{BH:92} Balbus, S.~A., \& Hawley, J.
~F. 1992, ApJ, 400, 595
\bibitem[Balbus \& Hawley (1998)]{BH:98} Balbus, S.~A., \& Hawley, J.~F. 1998, 
 Rev. Mod. Phys., 70, 1
\bibitem[Bardeen, Press, \& Teukolsky (1972)]{bpt72} 
Bardeen, J.~M., Press, W.~H. \& Teukolsky, S.~A. 1972, ApJ, 178, 347
\bibitem[De Villiers, \& Hawley (2002)]{DH:02} De Villiers, J.~P. \& 
 Hawley, J.~F. 2002, ApJ, 577, 866 (DH02)
\bibitem[De Villiers, \& Hawley (2003)]{DH:03} De Villiers, J.~P. \& 
 Hawley, J.~F. 2003, ApJ, in press (DH03)
\bibitem[Evans, \& Hawley (1988)]{EH:88} Evans, C.~R. \& Hawley, J.~F.
 1988, ApJ,  332, 659 
\bibitem[Gammie , McKinney \& Toth(2002)]{GM00} Gammie, C.~F.,  McKinney, J.,
\& T\'oth, G. 2003, ApJ, in press
\bibitem[Goodman \& Xu (1994)]{gx94} Goodman, J., \& Xu, G. 1994, 
ApJ, 432, 213
\bibitem[Hawley (2000)]{H00} Hawley, J.~F. 2000, ApJ, 528, 462 (H00)
\bibitem[Hawley (2001)]{H01} Hawley, J.~F. 2001, ApJ, 554, 534 
\bibitem[Hawley \& Balbus (1992)]{hb92} Hawley, J.~F., \& Balbus,
S.~A. 1992, ApJ, 400, 595
\bibitem[Hawley \& Balbus (2002)]{hb02} Hawley, J.~F., \& Balbus,
S.~A. 2002, ApJ, 573, 738
\bibitem[Hawley, Balbus, \& Stone (2001)]{hbs01} Hawley, J.~F., Balbus, S.~A., 
\& Stone, J.~M. 2001, ApJ, 554, L49
\bibitem[Hawley \& Krolik (2001)]{HK:01} 
 Hawley, J.~F. \& Krolik, J.~H., 2001, ApJ, 548, 348 
\bibitem[Hawley \& Krolik (2002)]{HK:02}
 Hawley, J.~F. \& Krolik, J.~H., 2002, ApJ, 566, 164
\bibitem[Hawley, Smarr \& Wilson (1984)]{HSW:84} 
 Hawley, J.~F., Smarr, L.~L., \& Wilson, J.~R., 1984, ApJ, 277, 296
\bibitem[Koide, Shibata, \& Kudoh (1999)]{ksk99} Koide, S., Shibata, K.,
\& Kudoh, T. 1999, ApJ, 522, 727
\bibitem[Krolik \& Hawley (2002)]{hk02} Krolik, J.~H., \& Hawley, J.~F. 2002,
ApJ, 573, 754
\bibitem[Machida \& Matsumoto (2003)]{mm03} Machida, M., \& Matsumoto, R. 
2003, ApJ, in press
\bibitem[Nelson, \& Papaloizou (2003)]{np03} Nelson, R., \&
Papaloizou, J.~C.~B. 2003, MNRAS, in press
\bibitem[Paczy\'nski \& Wiita 1980]{pw80}Paczy\'nski, B., \& Wiita, P.
J.  1980, A\&A, 88, 23
\bibitem[Papaloizou \& Pringle (1984)]{PP:84}
Papaloizou, J.~C.~B., \& Pringle, J.~E. 1984, MNRAS, 208, 721
\end{thebibliography}
\end{document}